\documentclass[]{aa}
\usepackage{graphics}

\begin {document}
\title{The NASA Astrophysics Data System: The Search
Engine and its User Interface}

\thesaurus{04(04.01.1)}
\author{G. Eichhorn\and M. J. Kurtz\and A. Accomazzi\and C. S. Grant
\and S. S. Murray}

\institute{Harvard-Smithsonian Center for Astrophysics, Cambridge, MA 02138}

\offprints{G. Eichhorn}
\mail{G. Eichhorn}

\date{Received / Accepted}

\titlerunning{}
\authorrunning{G. Eichhorn et al.}

\maketitle

\sloppy

\begin {abstract}
The ADS Abstract and Article Services provide access to the
astronomical literature through the World Wide Web (WWW).  The forms
based user interface provides access to sophisticated searching
capabilities that allow our users to find references in the fields of
Astronomy, Physics/Geophysics, and astronomical Instrumentation and
Engineering.  The returned information includes links to other on-line
information sources, creating an extensive astronomical digital
library.  Other interfaces to the ADS databases provide direct access
to the ADS data to allow developers of other data systems to integrate
our data into their system.

The search engine is a custom-built software system that is
specifically tailored to search astronomical references.  It includes
an extensive synonym list that contains discipline specific knowledge
about search term equivalences.

Search request logs show the usage pattern of the various search
system capabilities.  Access logs show the world-wide distribution of
ADS users.

The ADS can be accessed at http://adswww.harvard.edu

\end{abstract}

\section{Introduction}

The Astrophysics Data System (ADS) provides access to the astronomical
literature through the World Wide Web (WWW).  It is widely used in the
astronomical community.  It is accessible to anybody world-wide
through a forms based WWW interface.  A detailed description of the
history of the ADS is presented in the ADS Overview article
(\cite{adsoverview}, hereafter OVERVIEW).  The system contains
information from many sources (journals, other data centers,
individuals).  A detailed description of the data that we get and how
they are included in the ADS is presented in the ADS Data article
(\cite{adsdata}, hereafter DATA).  The incoming data are processed and
indexed with custom-built software to take advantage of specialized
knowledge of the data and the astronomical context.  A description of
this processing is given in the ADS Architecture article
(\cite{adsarchitecture}, hereafter ARCHITECTURE).  This article
describes the development and the current status of the ADS Abstract
Service user interface and search engine.

The ADS was created as a system to provide access to astronomical data
(\cite{1992ald2.proc..387M}).  In 1993 the ADS started to provide
access to a set of abstracts obtained from the NASA/STI (National
Aeronautics and Space Administration/Scientific and Technical
Information) project (\cite{1993adass...2..132K}).  The user interface
was built with the proprietary software system that the ADS used at
that time.  The search engine of this first implementation used a
commercial database system.  A description of the system at that time
is in \cite{1994ExA.....5..205E}.

In 1994, the World Wide Web (WWW, \cite{www}) became widely useful
through the NCSA Mosaic Web Browser (\cite{mosaic}).  The design of
the ADS Abstract Service with a clean separation between the user
interface and the search engine made it very easy to move the user
interface from the proprietary ADS system to the WWW.  In February
1994, a WWW interface to the ADS Abstract Service was made available
publicly.  The WWW interface to the ADS is described by
\cite{1995adass...4...28E} and \cite{1995VA.....39..217E}.  Within one
month of the introduction of the WWW interface, the usage of the
Abstract Service tripled, and it has continued to rise ever since
(\cite{1997Ap&SS.247..189E}).

With the increased usage of the system due to the easy access through
the WWW, severe limitations of the underlying commercial database
system very quickly became apparent.  We soon moved to an
implementation of the search engine that was custom-built and tailored
to the specific requirements of the data that we used.

In January 1995 we started to provide access to scanned journal
articles (\cite{1996adass...5..558A}).  The user interface to these
scans provided the user with the capability to access the scans in
various formats, both for viewing and for printing.

With time, other interfaces to the abstracts and scanned articles were
developed to provide other data systems the means to integrate ADS
data into their system (\cite{1996adass...5..569E}).

With the adoption of the WWW user interface and the development of the
custom-built search engine, the current version of the ADS Abstract
Service was basically in place.  The following sections describe the
current status of the different access capabilities
(sections~\ref{dataretrieval} and \ref{cookies}), the search engine
(sections~\ref{searchengine} and \ref{optimization}), access
statistics for the ADS system (section~\ref{accessstats}), and future
plans for the ADS interface and search engine (section~\ref{future}).

\section{\label {dataretrieval} Data Access}

The ADS services can be accessed through various interfaces.  Some of
these interfaces use WWW based forms, others allow direct access to
the database and search system through Application Program Interfaces
(APIs).  This section describes the various interfaces and their use,
as well as the returned results.

\subsection{Forms Based Interfaces}

\subsubsection{\label {abstracts} Abstract Service}

a. User Interface

\begin{figure}
\resizebox{\hsize}{!}{\includegraphics{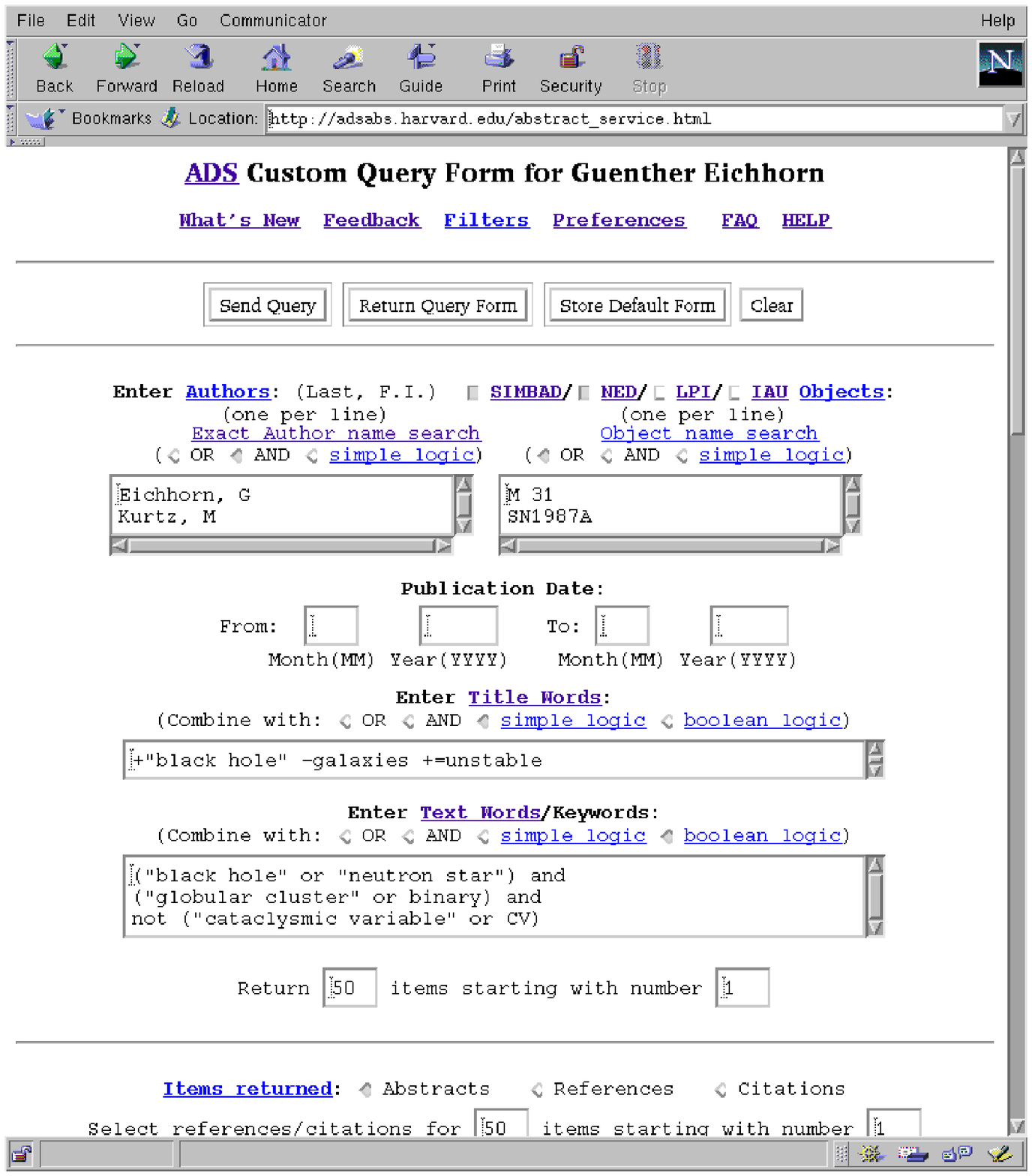}}
\caption[]{The ADS Abstract Service query form provides the capability to query the database by authors, object names, title and text words. }
\label{queryforma}
\end{figure}

\begin{figure}
\resizebox{\hsize}{!}{\includegraphics{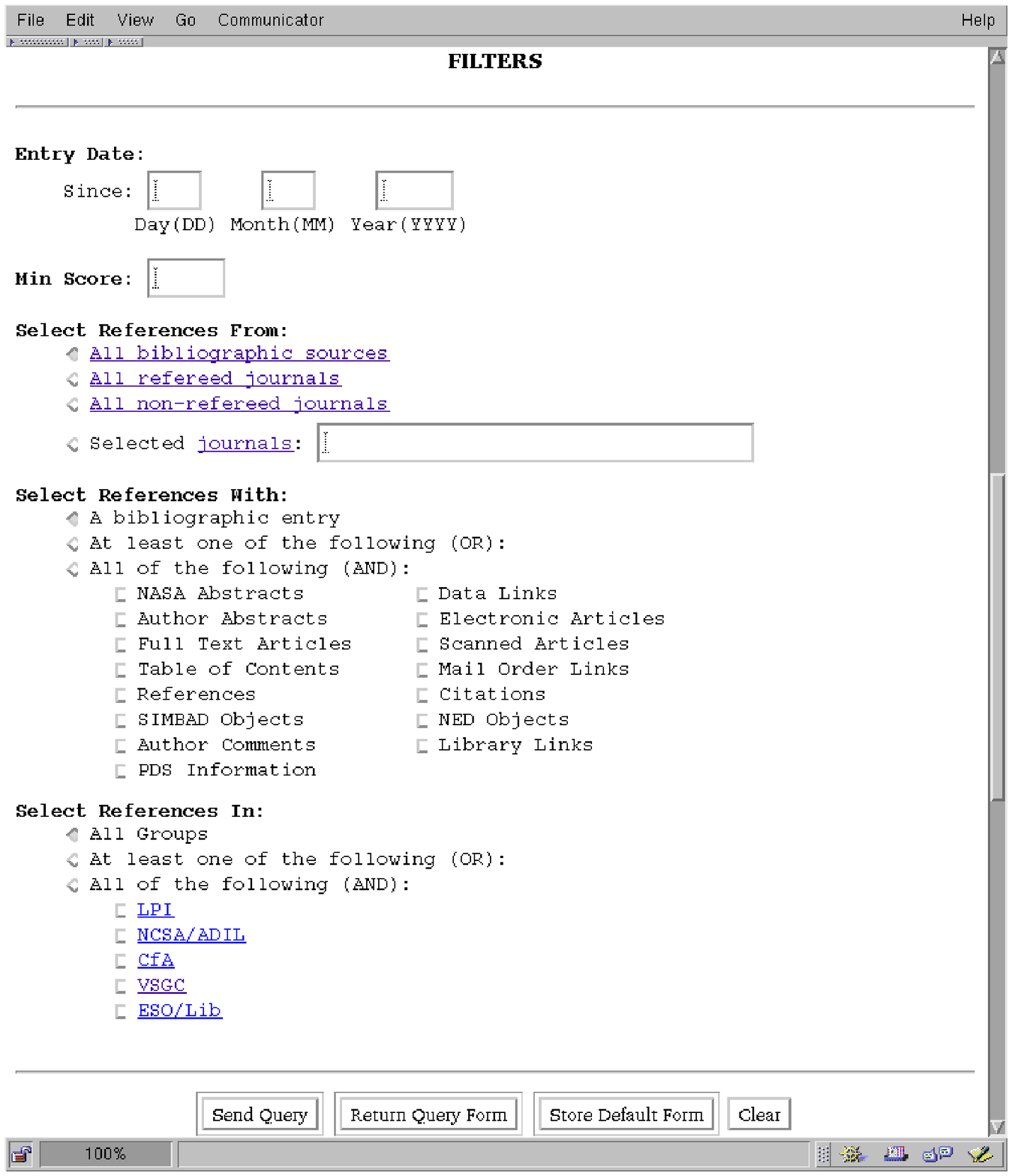}}
\caption[]{The Filter section of the query form allows selection of references that have specific properties. }
\label{queryformb}
\end{figure}

\begin{figure}
\resizebox{\hsize}{!}{\includegraphics{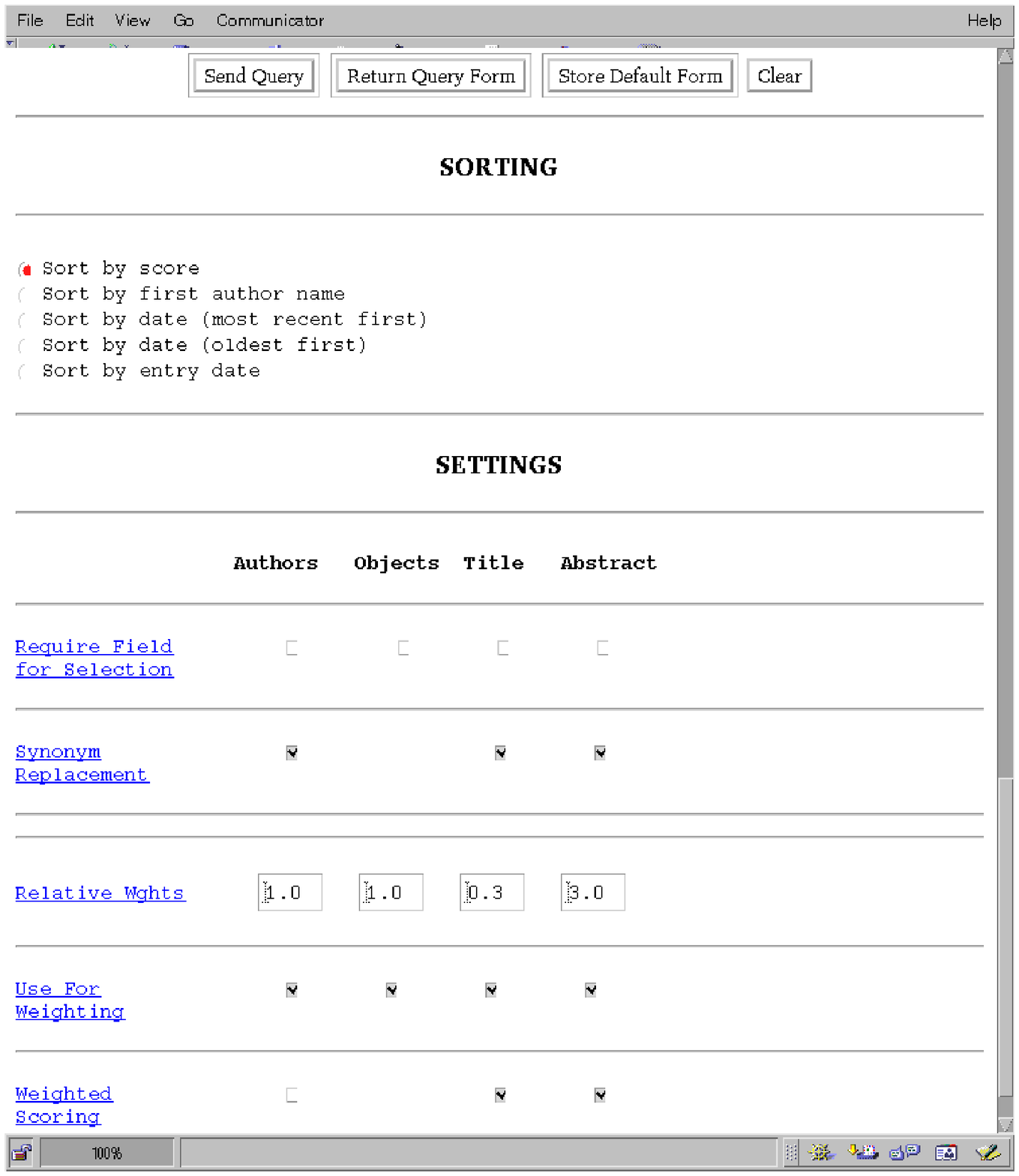}}
\caption[]{The Settings section of the query form allows the user to customize the search. }
\label{queryformc}
\end{figure}

The main query forms
(figures~\ref{queryforma},~\ref{queryformb},~\ref{queryformc}) provide
access to the different abstract databases.  These forms are generated
on demand by the ADS software.  This allows the software to check the
user identification through the HTTP (HyperText Transfer Protocol)
cookie mechanism (see section~\ref{cookies}), so that the software can
return a customized query form if one has been defined by the user.
It also adapts parts of the form according to the capabilities of the
user's web browser.

The query form allows the user to specify search terms in different
fields.  The input parameters in each query field can be combined in
different ways, as can the results obtained from the different fields
(figure~\ref{queryforma}).  The user can specify how the results are
combined through settings on the query form
(figure~\ref{queryformc}).  The combined results can then be filtered
according to various criteria (figure~\ref{queryformb}).

The database can be queried for author names, astronomical objects
names, title words, and words in the abstract text.  References can be
selected according to the publication date.  The author name, title,
and text fields are case insensitive.  The object field is case
sensitive when the IAU (International Astronomical Union) Circulars
(IAUC) object name database is searched, since the IAU object names
are case sensitive.  In the author and object name fields, the form
expects one search term per line since the terms can contain blanks.
In the title and text fields line breaks are not significant.

\medskip

\paragraph{Author Name Field}

The author names are indexed by last name and by a combination of last
name and first initial, separated by a comma.  To account for
differences in the spelling of the same author name, the search system
contains a list of author names that are spelled differently but are
in fact names of the same author.  This allows for automatically
retrieving all versions of common spelling differences.  This is
useful for instance for German umlaut spelled as Muller and Mueller,
or variations in the transliteration of names from non-English
alphabets like Cyrillic.  An example of such an entry in the author
synonym list is:

\begin{verbatim}
AFANASJEV, V
AFANAS'EV, V
AFANAS'IEV, V
AFANASEV, V
AFANASYEV, V
AFANS'IEV, V
AFANSEV, V
\end{verbatim}
Without this synonym replacement capability, author searches would
obviously be much less effective.  On user request we also include
name changes (e.g. due to marriage) in the author synonym list.
Combinations of search results within the author field use ``OR'',
``AND'', or simple logic (see below), depending on user selection.

Author names are quite often spelled differently in different
publications.  First names are sometimes spelled out, sometimes only
first initials are given, and sometimes middle initials are left out.
This makes it impossible to index all different spellings of a name
together automatically.  

To handle these different requirements, author names are indexed three
times, once with the last name only, once with the last name and first
initial, and once with the complete name as it is specified in the
article.

To access these different indexes, we provide two user interfaces for
author queries.  The regular user interface allows the user to search
for either a last name or a last name combined with the first initial.
This allows for fairly discriminating author searches.  It is a
compromise between the need to discriminate between different authors,
and the need to find all instances of a given author.  It
identifies all different versions of a given author quite reliably,
but it indexes together different authors with the same first initial.
For cases where this search method is not discriminating enough, we
provide a second user interface to the index of the full names, which
does not attempt to index different spellings of the same author
together.  When the user selects ``Exact Author Search'' and specifies
an author's last name or last name and first initial, a form is
returned with all distinct full author names that match the specified
name.  The user then selects all the different spellings of the
desired name and queries the database for articles that contain any
one of these different versions of an author's name.  For instance
specifying:
\begin{verbatim}
Eichhorn, G
\end{verbatim}

in the exact author name form returns the list:
\begin{verbatim}
EICHHORN, G.
EICHHORN, GERHARD
EICHHORN, GUENTHER
EICHHORN, GUNTHER
\end{verbatim}

Selecting the first, third, and fourth author name from that list will
return all articles by the first author of this article.  Any articles
by the second author containing only the first initial will also be
returned, but this is unavoidable.

\medskip

\paragraph{Object Name Field}

This field allows the user to query different databases for references
with different astronomical objects.  The databases that provide
object information are: SIMBAD (Set of Identifications, Measurements
and Bibliographies for Astronomical Data) at the Centre des Donn\'ees
Astronomique de Strasbourg (CDS), France (\cite{simbad}); the NASA
Extragalactic Database (NED) at the Infrared Processing and Analysis
Center (IPAC), Jet Propulsion Laboratory (JPL), Pasadena, CA
(\cite{1992adass...1...47M}); the IAU Circulars (IAUC) and the Minor
Planet Electronic Circulars (MPEC), both provided by the Central
Bureau for Astronomical Telegrams (CBAT) at the Harvard-Smithsonian
Center for Astrophysics in Cambridge, MA (\cite{1980CeMec..22...63M});
and a database with objects from publications from the Lunar and
Planetary Institute (LPI) in Houston (mainly Lunar sample numbers and
meteorite names).  The user can select which of these databases should
be queried.  If more than one database is searched, the results of
these queries are merged.  The LPI database does not have any entries
in common with the other databases.  The SIMBAD, NED, and IAUC
databases sometimes have information about the same objects.

\medskip

\paragraph{Title and Abstract Text Fields}

These fields query for words in the titles of articles or books, and
in the abstracts of articles or descriptions of books respectively.
The words from the title of each reference are also indexed in the
text field so they will be found through either a title or a text
search.  Before querying the database the input in these fields is
processed as follows:

1. Apply translation rules.  This step merges common expressions into
a single word so that they are searched as one expression.  Regular
expression matching is used to convert the input into a standard
format that is used to search the database.  For instance {\it M 31}
(with a space) is translated to {\it M31} (without a space) for
searching as one search term.  In order to make this general
translation, a regular expression matching and substitution is
performed that translates all instances of an `M' followed by one or
more spaces or a hyphen followed by a number into `M' directly
followed by the number.  Other translation rules include the
conversion of {\it NGC 1234} to {\it NGC1234}, contractions of {\it T
Tauri, Be Star, Shoemaker Levy}, and several others (see ARCHITECTURE).

2. Remove punctuation.  In this step all non-alphanumeric characters
are removed, unless they are significant (for instance symbols used in the
simple logic (see below), `+' and `--' before numbers, or `.' within numbers).

3. Translate to uppercase.  All information in the index files is in
uppercase, except for object names from the IAU Circulars.

4. Remove kill words.  This step removes all non-significant words.
This includes words like `and', `although', `available', etc (for more
details see ARCHITECTURE).

In the title and text fields, searching for phrases can be specified
by enclosing several words in either single or double quotes, or
concatenating them with periods (`.') or hyphens (`--').  All these
accomplish the same goal of searching the database for references that
contain specified sequences of words.  The database is indexed for
two-word phrases in addition to single words.  Phrases with more than
two words are treated as a search for sets of two-word phrases
containing the first and second word in the first phrase, the second and
third word in the second phrase, etc.

\medskip

b. Searching

After the search terms are pre-processed, the databases of the
different fields are searched for the resulting list of words, the
results are combined according to the selected combination rules, and
the resulting score is calculated according to the selected scoring
criteria.  These combination rules provide the means for improving the
selectivity of a query.

\medskip

\paragraph{Search Word Selection}

The database is searched for the specified words as well as for words
that are synonymous with the specified term.  One crucial part to
successful searches in a free text search system is the ability to not
only find words exactly as specified, but also similar words.  This
starts with simply finding singular and plural forms of a word, but
then needs to be extended to different words with the same meaning in
the normal usage of words in a particular field of science.  In
Astronomy for instance ``spectrograph'' and ``spectroscope'' have
basically the same meaning and both need to be found when one of these
words is specified in the query.  Even further reaching, more
discipline-specific synonyms are necessary for efficient searches such
as ``metallicity'' and ``abundance'' which have the same meaning in
astronomical word usage.  In order to exhaustively search the database
for a given term, it is important to search for all synonyms of a
given word.  The list of synonyms was developed manually by going
through the list of words in the database and grouping them according
to similar meanings.  This synonym list is a very important part of
the ADS search system and is constantly being improved (see
ARCHITECTURE).

The list of synonyms also contains non-English words associated with
their English translations.  These words came from non-English
reference titles that we included in the database.  This allows searches
with either the English or non-English words to find references with
either the English word or the non-English translation.  We are in the
process of extending this capability by including translations of most
of the words in our database into several languages (German, French,
Italian, Spanish).  This will allow our users to phrase queries in any
of these languages.  We expect to complete this project sometime in 2000.

By default a search will return references that contain the search
word or any of its synonyms.  The user can choose to disable this
feature if for some reason a specific word needs to be found.  The
synonym replacement can be turned off completely for a field in the
``Settings'' section of the query form.  This can be used to find a
rare word that is a synonym of a much more frequent word, for
instance if you want to look for references to ``dateline'', which is a
synonym to ``date''.  Synonym replacement can also be enabled or
disabled for individual words by prefixing a word with `=' to force an
exact match without synonym replacement.  When synonym replacement is
disabled for a field, it can be turned on for a particular word by
prefixing it with `\#'.

\paragraph{Selection Logic Within a Field}

There are four different types of combinations of results for searches
within a field possible.
\begin{verbatim}
1.  OR
2.  AND
3.  Simple logic
4.  Full boolean logic
\end{verbatim}

1. Combination by `OR':
The resulting list contains all references that contain at least one
of the search terms.

2. Combination by `AND':
The resulting list has only references that contain every one
of the search terms.

3. Combination by simple logic:
The default combination in this logic
is by `OR'.  Individual terms can be either required for selection by
prefixing them with a `+', or can be selected against by prefixing
them with a `--'.  In the latter case only references that do not
contain the search term are returned.  If any of the terms in the
search is prefixed by a `+', any other word without a prefix does not
influence the resulting list of references.  However, the final score
(see below) for each reference will depend on whether the other search
terms are present.

4. Combination by full boolean logic:
In this setting, the user specifies a boolean expression containing
the search terms and the boolean operators `and', `or', and `not', as
well as parentheses for grouping.  A boolean expression could for
instance look like:

(pulsar or ``neutron star'') and (``red shift'' distance) and not 1987A

This expression searches for references that contain either the word
pulsar or the phrase ``neutron star'' and either the phrase ``red
shift'' or the word distance (``or'' being the default), but not the
word 1987A.

\medskip

\paragraph{Selection Logic Between Fields}

In the settings part of the query form, the user can specify fields
that will be required for selection.  If a field is selected as
``Required for Selection'' only references that were selected in the
search specified in that field will be returned.  If one field is
selected as ``Required for Selection'', the searches in fields that are
not set as ``Required for Selection'' do not influence the resulting
list, but they influence the final score.

\medskip

c. Scoring

The list of references resulting from a query is sorted according to a
``score'' for each reference.  This score is calculated according to how
many of the search items were matched.  The user has the choice
between two scoring algorithms:
\begin{verbatim}
1.  proportional scoring
2.  weighted scoring
\end{verbatim}
These scoring algorithms have been analyzed by \cite{scoring}.

In proportional scoring, the score is directly proportional to the
number of terms found in the reference.  In weighted scoring, the
score is proportional to the inverse logarithm of the frequency of the
matched word.  This weighting gives higher scores for words that are
less frequent in the database and therefore presumably more important
indicators of the relevance of a match.  In the settings section of
the query form the user can select which type of scoring should be
used for each query field separately.  The default setting for title
and text searches is the weighted scoring.  For author searches
proportional scoring is the default.  Once the score for each query
field is calculated, the scores are normalized so that a reference
that matches all words in a field receives a score of 1.

The normalized scores from the different fields are then combined to
calculate a total score.  Again the result is normalized so that a reference
that matches all words in each query field has a score of 1.  The user
can influence this combining of scores from the different search
fields by assigning weights to the different fields.  This allows the
user to put more emphasis in the selection process on, for instance, the
object field by assigning a higher weight to that field.  Another
use of the weight field is to select against a field.  For instance
specifying an object name and an author name and selecting a negative
weight to the author field will select articles about that object that
were {\em not} written by the specified author.

The relative weights for the different search fields can be set by the
user.  The ADS provides default weights as follows:
\begin{verbatim}
Authors: 1.0
Objects: 1.0
Title:   0.3
Text:    3.0
\end{verbatim}

These default weights were determined on theoretical grounds, combined
with trial and error experimentation.  We used different search inputs
from known research fields and different weights and ranked the
resulting lists according to how well they represented articles from
these research fields.  The weights listed above gave the best
results.

\medskip

d. Filtering of Selected References

The selected references can be filtered according to different
criteria (see section~\ref{filters}) in order to reduce the number of
returned references.  The user can select references according to
their entry date in the database, a minimum score (see above), the
journal they are published in, whether they have pointers to selected
external data sources, or whether they belong to one or more of
several groups of references.  This allows a user for instance to
select only references from refereed journals or from one particular
journal by specifying its abbreviation.  It also allows a user to
select only references that have links to external data sets, on-line
articles, or that have been scanned and are available through the ADS
Article Service.

\medskip

e. Display of Search Results

The ADS system returns different amounts of information about a
reference, depending on what the user request was.  This section
describes the different reference formats.

\medskip

\paragraph{Short Reference Display}

The list of references returned from a query is displayed in a
tabular format.  The returned references are sorted by score first.
For equal scores, the references are sorted by publication date with
the latest publications displayed first.

\begin{figure}
\resizebox{\hsize}{!}{\includegraphics{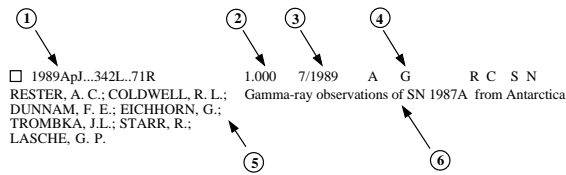}}
\caption[]{Elements in the references returned from an ADS query. }
\label{shortelements}
\end{figure}

A typical reference display is shown in figure~\ref{referencedisp}.  The
fields in such a reference are shown in figure~\ref{shortelements}.
They are as follows:

\begin{figure}
\resizebox{\hsize}{!}{\includegraphics{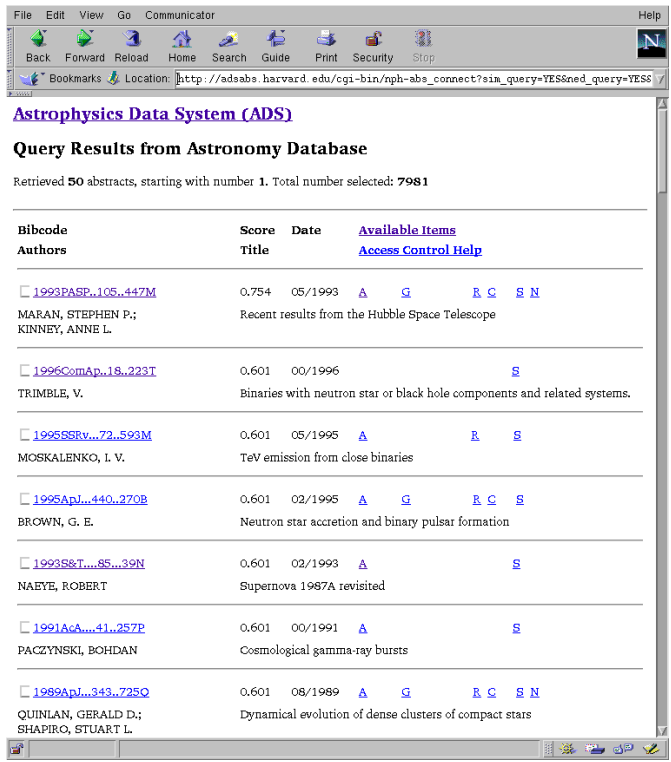}}
\caption[]{Entries in the list of references returned by an ADS query contain the bibliographic code (1) the matching score (2), the publication date (3), a list of data links (4), the list of authors (5), and the title of the reference (6). }
\label{referencedisp}
\end{figure}

1.  Bibliographic Code: This code identifies the reference uniquely
(see DATA and \cite{1995ioda.book..259S}).
Two important properties of these codes are that they can be generated
from a regular journal reference, and that they are human readable and
can be understood and interpreted.

2.  Score:  The score is determined during the search according to
how well each reference fits the query.

3.  Date:  The publication date of the reference is displayed as mm/yyyy.

4.  Links: The links are an extremely important aspect of the ADS.  They
provide access to information correlated with the article.
Table~\ref{linkletters} shows the links that we currently provide
when available.

\begin{table*}
\caption[]{Links types and their numbers in the ADS database.
}
\label{linkletters}
\begin{tabular*}{7.0in}{llp{0.7\linewidth}}

\hline
\noalign{\smallskip}

Link & Resource & Description

\\
\noalign{\smallskip}
\hline
\noalign{\smallskip}
A & Abstract & Full abstract of the article.  These abstracts come
                    from different sources.\\

C & Citations & A list of articles that cite the current article.
                    This list is not necessarily complete (see `R'
                    References).\\

D & On-line Data & Links to on-line data at other data centers.\\

E & Electronic Article & Links to the on-line version of the article.
                    These on-line versions are in HTML format for
                    viewing on-screen, not for
                    printing.$^{\mathrm{a}}$\\

F & Printable Article & Links to on-line articles in PDF or
                    Postscript format for
                    printing.$^{\mathrm{a}}$\\
                     
G & Gif Images & Links to the images of scanned articles in the ADS
                    Article Service.\\

I & Author Comments & Links to author supplied additional
					information (e.g. corrections, additional
	                references, links to data),\\

L & Library Entries & Links to entries in the Library of Congress
                    on-line system.\\

M & Mail Order & Links to on-line document delivery systems at the
                    publisher/owner of the article.\\

N & NED Objects & Access to lists of objects for the current article
                    in the NED database.\\

O & Associated Articles & A list of articles that are associated
                    with the current article.  These can be errata or
                    other articles in a series.\\
P & Planetary Data System & Links to datasets at the Planetary Data
					System.\\

R & References & A list of articles referred to in the current
                    article.  For older articles these lists are not
                    necessarily complete, they contain only references
                    to articles that are in the ADS database.  For
                    articles that are on-line in electronic form, the
                    `R' link points to the on-line reference list, and
                    therefore the complete list of references in that
                    article.$^{\mathrm{a}}$\\

S & SIMBAD Objects & Access to lists of objects for the current
                    article in the SIMBAD database.\\

T & Table of Contents & Links to the list of articles in a books or
					proceedings volume.\\

X & Planetary Nebulae & Links to datasets in the Galactic Planetary
					Nebulae Database.

\\
\noalign{\smallskip}
\hline
\end{tabular*}
\begin{list}{}{}
\item[$^{\mathrm{a}}$]There
is generally access control at the site that serves these on-line articles
\end{list}{}{}
\end{table*}

A more detailed description of resources in the ADS that these links
point to is provided in DATA.

Some of these links (for instance the `D' links) can point to more
than one external information provider.  In such cases the link points
to a page that lists the available choices of data sources.  The user
can then select the more convenient site for that resource, depending
on the connectivity between the user site and the data site.

5.  Authors: This is the list of authors for the reference.  Generally
these lists are complete.  For some of the older abstracts that we
received from NASA/STI, the author lists were truncated at 5 or 10
authors, but every effort has been made to correct these abbreviated
author lists (see DATA).

6.  Title: The complete title of the reference.

The reference lists are returned as forms if table display is selected
(see section~\ref{cookies}).  The user can select some or all of the
references from that list to be returned in any one of several formats:

i.  HTML format: The HTML (HyperText Markup Language) format is for
     screen viewing of the formatted record.

ii.  Portable Format:  This is the format that the ADS uses internally
     and for exchanging references with other data centers.  A
     description of this format is available on-line at:
\begin{verbatim}
http://adsabs.harvard.edu/abs_doc/
       abstract_format.html
\end{verbatim}

iii. BibTeX format: This is a standard format that is used to build
     reference lists for TeX (a typesetting language especially suited
     for mathematical formulas) formatted articles.

iv.  ASCII format:  This is a straight ASCII text version of the
     abstract.  All formatting is done with spaces, not with tabs.

v.   User Specified Format:  This allows the user to specify in which
     format to return the reference.  The default format for this
     selection is the bibitem format from the AASTeX macro package.
     The user can specify an often used format string in the user
     preferences (see section~\ref{cookies}).  This format string
     will then be used as the default in future queries.

The user can select whether to return the selected abstracts to the
browser, a printer, a local file for storage, or email it to a
specified address.

\medskip

\paragraph{Full Abstract Display}

\begin{figure}
\resizebox{\hsize}{!}{\includegraphics{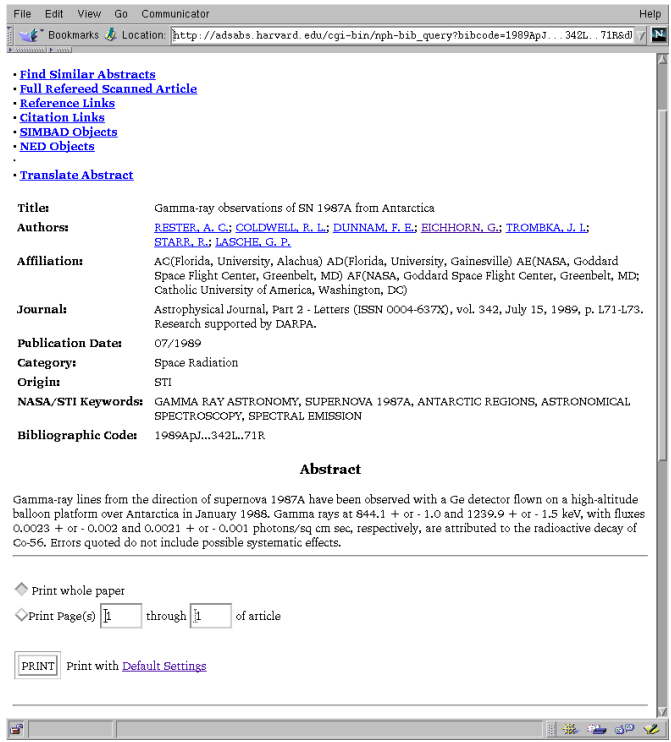}}
\caption[]{The full abstract display contains (where available) the title, author list, journal information, author affiliations, publication date, keywords, the origin of the reference, the bibliographic code, the abstract, object names, abstract category, and a copyright notice. }
\label{abstract}
\end{figure}

In addition to the information in the short reference list, the full
abstract display (see figure~\ref{abstract}) includes, where available,
the journal information, author affiliations, language, objects, keywords,
abstract category, comments, origin of the reference, a copyright
notice, and the full abstract.  It also includes all the links
described above.

For abstracts that are displayed as a result of a search, the system
will highlight all search terms that are present in the returned
abstract.  This makes it easy to locate the relevant parts in an
abstract.  Since the highlighting is somewhat resource intensive, it
can be turned off in the user preference settings (see
section~\ref{cookies}).

For convenience, the returned abstract contains links that allow the
user to directly retrieve the BibTex or the custom formatted version of
the abstract.

The full abstract display also includes a form that provides the
capability to use selected information from the reference to build a
new query to find similar abstracts.  The query feedback mechanism
makes in-depth literature searches quick and easy.  The user can
select which parts of the reference to use for the feedback query
(e.g. authors, title, or abstract).  The feedback query can either be
executed directly, or be returned as a query form for further
modification before executing it, for instance to change the
publication date range or limit the search to specific journals.  This
query feedback mechanism is a very powerful means to do exhaustive
literature searches and distinguishes the ADS system from most other
search systems.  A query feedback ranks the database against the
record used for the feedback and sorts it according to how relevant
each reference is to the search record.  The query feedback can be
done across databases.  For instance a reference from the Astronomy
database can be used as query feedback in the preprint database to see
the latest work in the field of this article.

If the article for the current reference has been scanned and is available
through the ADS Article Service (see below), printing options are
available in the abstract display as well.  These printing options
allow the printing of the article without having to retrieve the
article in the viewer first.

\subsubsection{Article Service}

This part of the ADS provides access to the scanned images of
articles.  We have received permission from most astronomy journals to
scan their volumes and make them available on-line free of charge.  A
more detailed description of these data is in the DATA.

The most common access to the scanned articles is through the ADS
Abstract Service via the `G'-links (see above).  However they can also
be accessed directly through the article query page by
publication year and month or by volume and page at:

\begin{verbatim}
http://adsabs.harvard.edu/article_service.html
\end{verbatim}

This form returns the specified article in the user specified format
(see section~\ref{cookies}).  If a page within an article is
specified, and the single page display is selected, the specified page
within the article is returned with the links to the other available pages
as usual.  

\begin{figure}
\resizebox{\hsize}{!}{\includegraphics{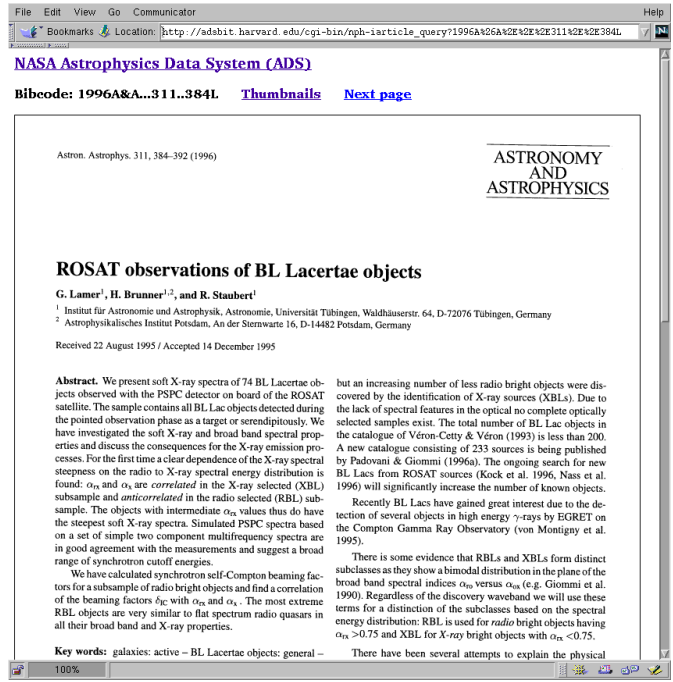}}
\caption[]{The article display shows a gif image of the selected journal page with the resolution selected in the user preferences. }
\label{articlegif}
\end{figure}

The article display normally shows the first page
(figure~\ref{articlegif}) of an article at the selected resolution and
quality (see section~\ref{cookies}).  The user can select resolutions
of 75, 100, or 150 dots per inch (dpi) and image qualities of 1, 2, 3,
or 4 bits of greyscale per pixel.  These gif images are produced on
demand from the stored tiff images (see DATA).  The default version of
the gif images (100 dpi, 3 bit greyscale) is cached on disk.  The
cache of these gif images is managed to stay below a maximum size.
Any time the size of the cached gif images exceeds the preset cache
size, the gif images of pages that have not been accessed recently
are deleted.

Below the page image on the returned page are links to every page of
the article individually.  This allows the user to directly access any
page in the article.  Wherever possible, plates that have been printed
separately in the back of the journal volume have been bundled with
the articles to which they belong for ease of access.  The next part
of the displayed document provides access to plates in that volume
if the plates for this journal are separate from the articles.
Another link retrieves the abstract for this article.

The next part of the page allows the printing of the article.  If the
browser works with HTTP persistent cookies (see
section~\ref{cookies}), there is just one print button in that section
with a selection to print either the whole paper or individual pages.
This print button will print the article in the format that the user
has specified in the user preferences.  If the browser does not handle
cookies, several of the more commonly used print options are made
available here.

All possible printing options can be accessed through the next link
called ``More Article Retrieval Options''.  This page allows the user to
select all possible retrieval options.  These include:

i.  Postscript: Access to two resolutions is provided (200 dpi and 600
dpi).  For compatibility with older printers, there is also an option
to retrieve Postscript Level 1 files.

ii.  PCL (Printer Control Language): This language is for printing on
PCL printers such as the HP desk jets and compatibles.

iii. PDF (Portable Document Format): PDF can be viewed with the Adobe
Acrobat reader (\cite{acrobat}).  From the Acrobat reader the article
can be printed.

iv.  TIFF (Tagged Image File Format): The original images can be
retrieved for local storage.  This would allow further processing like
extraction of figures, or Optical Character Recognition (OCR) in order
to translate the article into ASCII text.

v.   Fax retrieval:  Within the USA, articles can be retrieved via
fax at no cost.  Again, the retrieval is greatly facilitated through
the preferences setting capability.  The preferences allow the user to
store a fax number that will be used for the fax service.

vi.  Email retrieval: Articles can be retrieved through email instead
of through a WWW browser.  MIME (Multipurpose Internet Mail Extension,
\cite{mime}) capable email systems should be able to send the
retrieved images directly to the printer, to a file, or to a viewer,
depending on what retrieval option was selected by the user.

\medskip

For most of the retrieval options, the data can optionally be compressed
before they are sent to the user.  Unix compress and GNU gzip are
supported compression algorithms.

Instead of displaying the first page of an article together with the
other retrieval links, the user has the option (selected through the
preferences system, see section~\ref{cookies}) to display thumbnails
of all article pages simultaneously.  This allows an overview of the
whole article at once.  One can easily find specific figures or
sections within an article without having to download every page.
This should be especially useful for users with slow connections to
the Internet.  Each thumbnail image ranges in size from only 700 bytes
to 3000 bytes, depending on the user selected thumbnail image quality.
The rest of this type of article page is the same as for the
page-by-page display option.

\subsubsection{Other Forms Based User Interfaces}

There are several forms available to directly access references or
articles and other relevant information.  All abstract query forms
return the short reference format as described above.  One form allows
access to references through bibliographic codes:

\begin{verbatim}
http://adsabs.harvard.edu/bib_abs.html
\end{verbatim}

This form allows the user to retrieve abstracts by specifying directly
a bibliographic code or the individual parts of a bibliographic (year,
journal, volume, page).  This can be very useful in retrieving
references from article reference lists, since such reference lists
generally contain enough information to build the bibliographic codes
for the references.  This form also accepts partial codes and returns
all references that match the partial code.  It accepts the wildcard
character `?'.  The `?' wildcard stands for one character in the code.
For partial codes that are shorter than 19 characters, matching is
done on the first part of the bibliographic codes.  For instance:

1989ApJ...341?...1
\noindent
will retrieve the articles on page 1 of the ApJ (Astrophysical
Journal) and ApJ Letters volume 341, regardless of the author name.

Another form allows access to the Tables of Contents (ToCs) of selected
journals by month/year or volume:

\begin{verbatim}
http://adsabs.harvard.edu/toc_service.html
\end{verbatim}

One option on that form is to retrieve the latest published issue of a
particular journal.  Access to the last volumes of a set of journals
is also available though a page with a graphical display of selected
journals' cover pages:

\begin{verbatim}
http://adsabs.harvard.edu/tocs.html
\end{verbatim}

By clicking on a journal cover page either the last published volume
of that journal or the last volume that the user has not yet read is
returned, depending on the user preference settings (see
section~\ref{cookies}).  The information necessary for that service is
stored with the user preferences in our internal user preferences
database.

A customized ToC query page is available at:

\begin{verbatim}
http://adsabs.harvard.edu/custom_toc.html
\end{verbatim}

It will display only icons for journals that have issues available
that have not been read by the user.  This allows a user to see at a
glance which new issues for this set of journals have been published.
The set of journals that is included in the customized ToC query page
can be specified in the user preferences (see section~\ref{cookies}).

As mentioned in section~\ref{abstracts} and in ARCHITECTURE, one
important aspect of the ADS search system is the list of synonyms.
Sometimes it is important for our users to be able to see what words
are in a particular synonym group to properly interpret the search
results.  Another question that is asked is what words are in the
database and how often.  The list query page (linked to the words
``Authors'', ``Title Words'', and ``Text Words'' above the
corresponding entry fields on the main query form) allows the user to
find synonym groups and words in the database.  The user can specify
either a complete word in order to find its synonyms, or a partial
word with wildcard characters to find all matching words in the
database.  When a word without wildcard characters is specified, the
list query form returns all of its synonyms (if any).

To find words matching a given pattern, the users can specify a
partial word with either or both of two wildcard characters.  The
question mark `?'  stands for any single character, the asterisk `*'
stands for zero or more characters (see section~\ref{wildcard}).  For a
wildcard search, the list query form returns all words in the database
that match the specified pattern, together with the frequencies of
these words in the database.

\subsubsection{Journal specific access forms}

The regular query forms search the complete database.  The user can
select the return of only specific journals in the ``Filter'' section
of the query form.  In order to allow different journals to use the
ADS system for searching their references, journal specific pages are
available.  The URL (Uniform Resource Locator) for an abstract search
page for a specific journal is:

\begin{verbatim}
http://adsabs.harvard.edu/Journals/
       search/bibstem
\end{verbatim}

The page for retrieving scanned articles of a specific journal is:

\begin{verbatim}
http://adsabs.harvard.edu/Journals/
       articles/bibstem
\end{verbatim}

and the page for retrieving the tables of contents by volume or
publication date is:

\begin{verbatim}
http://adsabs.harvard.edu/Journals/toc/bibstem
\end{verbatim}

In each case, bibstem is the abbreviation for the selected journal.
For instance:

\begin{verbatim}
http://adsabs.harvard.edu/Journals/search/ApJ
\end{verbatim}

returns a query form for searching only references of the Astrophysical
Journal.  These forms are available for linking by anybody.

\subsection{Direct Access Interfaces}

Both abstracts and articles can be accessed directly though HTML
hyperlinks.  The references are identified through the bibliographic
codes (or bibcodes for short) mentioned above and described in detail
in DATA.  The syntax for such links to access
abstracts is:

\begin{verbatim}
http://adsabs.harvard.edu/cgi-bin/
       bib_query?bibcode=1989ApJ...342L..71R
\end{verbatim}

Scanned articles can be accessed directly through links of the form:

\begin{verbatim}
http://adsabs.harvard.edu/cgi-bin/
       article_query?bibcode=1989ApJ...342L..71R
\end{verbatim}

These links will return the abstract or scanned article respectively
for the specified bibliographic code.  They are guaranteed to
work in this form.  You may see other URLs while you use the ADS.
These are internal addresses that are not guaranteed to work in the
future.  They may change names or parameters.  Please use only links
of the form described above to directly access abstracts and articles.

\subsection{Embedded Queries}

Embedded queries can be used to build hyperlinks that return the
results of a pre-formulated query.  One frequently used example is a
link that returns all articles written by a specific user.  The syntax
for such a link is:

\begin{verbatim}
<a href="http://adsabs.harvard.edu/cgi-bin/
       abs_connect?return_req=no_params&
       param1=val1&param2=val2&...">...</a>
\end{verbatim}

There are no spaces allowed in a URL.  All blanks need to be encoded
as `+'.  The parameter return\_req=no\_params sets all the default
settings.  Individual settings can be changed by including the name of
the specific setting and its value after the return\_req=no\_params
parameter in the URL.  A list of available parameters can be accessed
at:

\begin{verbatim}
http://adsdoc.harvard.edu/cgi-bin/
       get_field_names.pl
\end{verbatim}

We try to make changes to parameters backward compatible, but that is
not always possible.  We encourage you to use this capability, but it
is advisable to use only the more basic parameters.

This type of interface allows users to link to the ADS for a
comprehensive list of references on a specific topic.  Many users use
this to provide an up-to-date publication list for themselves by
encoding an author query into an embedded query.

\subsection{\label{perlaccess} Perl Script Queries}

The ADS database can be used by other systems to include ADS data in
documents returned from that site.  This allows programmers at other
sites to dynamically include the latest available information in their
pages.  An example is the interface that SPIE (the International
Society for Optical Engineering) provides to our database.  It is
available at:

\begin{verbatim}
http://www.spie.org/incite/
\end{verbatim}

This site uses Perl scripts to query our database and format the
results according to their conventions.  These Perl scripts are
available at:

\begin{verbatim}
http://adsabs.harvard.edu/www/adswww-lib/
\end{verbatim}

The Perl scripts allow the programmer to specify all the regular
parameters.  The results are returned in Perl arrays.
If you use these scripts, we would appreciate it if you would credit
the ADS somewhere on your pages.

\subsection{\label{email} Email Interface}

The ADS Abstract Service can be accessed through an email interface.
This service is somewhat difficult to use since it involves an
interface between two relatively incompatible interface paradigms.
This makes describing it quite difficult as well.  It is intended for
users who do not have access to web browsers.  If you have questions
about how to use this access, please contact the ADS at
ads@cfa.harvard.edu.

To get information about this capability, send email to:

adsquery@cfa.harvard.edu

with the word ``help'' in the message body.

This interface accepts email messages with commands in the message
body.  The subject line is ignored.  The commands that are available
are:
\begin{verbatim}
1) help (see above)
2) action=URL
\end{verbatim}

The second command allows a user to retrieve a document at the
specified URL.  Three qualifiers allow the user to specify what
retrieval method to use, what format to return, and to which address
to return the results:
\begin{verbatim}
a) method=`method'
b) return=`return-type'
c) address=`e-mail-address'
\end{verbatim}

a. `method' is either `get' or `post' (without the quotes).  This
determines what kind of query will be executed.  To retrieve a form
for further queries, use the `get' method.  To execute a forms query
you need to know what type of query the server can handle.  If you
execute a forms query after retrieving the form through this service,
the correct method line will already be in place.  Default method is
`get'.

b. `return-type' is either `text', `form', or `raw' (without the
quotes).  If text return is requested, only the text of the query
result is returned, formatted as if viewed by a WWW browser.  If form
return is requested, the text of the result is returned as well as a
template of the form that can again be executed with this service.  If
raw return is requested, the original document is returned without any
processing.  Default return is `form'.  The capability to return MIME
encoded files is in preparation.

c. `e-mail-address' specifies to which e-mail address the result
should be sent.  This line is optional.  If no address is specified,
the result is sent to the address from where the request came.

To execute a query via email, the user first retrieves the abstract
query form with:

\begin{verbatim}
action=http://adsabs.harvard.edu/
              abstract_service.html
\end{verbatim}

This will return an executable form.  This form can be returned to the
ADS in an email message to execute the query.  The user enters input
for the different fields as required in the forms template.  For forms
tags like checkboxes or radio buttons, the user can uncomment the
appropriate line in the form.  Comments in the form that are included
with each forms field provide guidance for completing the form before
submission.  The text part of the form is formatted as comment lines
so that the user does not have to modify any irrelevant parts of the
form.  The retrieval method is already set appropriately.

\subsection{Z39.50 Interface}

\fussy
The recently implemented Z39.50 interface (\cite{z3950}) conforms to
the Library of Congress Z39.50 conventions
(http://lcweb.loc.gov/z3950/lcserver.html).  This allows any library
that uses this protocol to access the ADS through their interface.
The databases supported through this interface are listed in
table~\ref{z3950db}, search fields supported are listed in
table~\ref{z3950use}, supported relationship attributes are listed in
table~\ref{z3950rel} and the supported structure attributes are listed
in table~\ref{z3950strct}.  Table~\ref{z3950recfmt} lists the
supported record formats and table~\ref{z3950recsyn} shows the
supported record syntax.
\sloppy

\begin{table}
\caption[]{Database Names Supported in the ADS database. }
\label{z3950db}
\begin{tabular*}{3.4in}{lp{0.8\linewidth}}

\hline
\noalign{\smallskip}

Value & Description\\

\noalign{\smallskip}
\hline
\noalign{\smallskip}
AST & Astronomy Database.  Contains references from
      Astronomy related articles\\
INST & Instrumentation Database.  Contains references from
       articles related to Space Instrumentation\\
PHY & Physics Database.  Contains references from
      Physics related articles\\
PRE & Preprint Database.  Contains references from
      the Los Alamos preprint server related to Astronomy\\
GEN & General Database.  Contains references from
      articles unrelated to the other databases

\\
\noalign{\smallskip}
\hline
\end{tabular*}
\end{table}

\begin{table}
\caption[]{Use Attributes Supported in the ADS database. }
\label{z3950use}
\begin{tabular*}{3.4in}{ll}

\hline
\noalign{\smallskip}

Value & Description

\\
\noalign{\smallskip}
\hline
\noalign{\smallskip}
1 & Personal name$^{\mathrm{a}}$\\
1003 & Author$^{\mathrm{a}}$\\
4 & Title\\
5 & Title series$^{\mathrm{b}}$\\
62 & Abstract\\
31 & Publication Date\\
1011 & Entry Date in Database

\\
\noalign{\smallskip}
\hline
\end{tabular*}
\begin{list}{}{}
\item[$^{\mathrm{a}}$]These attributes search the same field
\item[$^{\mathrm{b}}$]This attribute limits searches to the journal specified
\end{list}{}{}
\end{table}

\begin{table}
\caption[]{Relation Attributes Supported in the ADS database.
}
\label{z3950rel}
\begin{tabular*}{3.4in}{llp{0.4\linewidth}}

\hline
\noalign{\smallskip}

Value & Description & Fields Supporting

\\
\noalign{\smallskip}
\hline
\noalign{\smallskip}
3 & Equal$^{\mathrm{a}}$ & All\\
2 & Less than or Equal & Publication Date\\
4 & Greater than or Equal & Publication Date, Entry Date\\
102 & Relevance$^{\mathrm{b}}$ & Title, Abstract, Author

\\
\noalign{\smallskip}
\hline
\end{tabular*}
\begin{list}{}{}
\item[$^{\mathrm{a}}$]Relation Equal searches for exact words.
\item[$^{\mathrm{b}}$]Relation Relevance searches for words and their synonyms.
\end{list}{}{}
\end{table}

\begin{table}
\caption[]{Structure Attributes Supported in the ADS database.
}
\label{z3950strct}
\begin{tabular*}{3.4in}{lll}

\hline
\noalign{\smallskip}

Value & Description & Fields Supporting

\\
\noalign{\smallskip}
\hline
\noalign{\smallskip}
1 & quoted phrase & Title, Abstract\\
2 & word & Title, Abstract, Author, Series\\
6 & word list & Title, Abstract, Author, Series\\
5 & date & Publication Date, Entry Date

\\
\noalign{\smallskip}
\hline
\end{tabular*}
\end{table}

\begin{table}
\caption[]{Record Format Supported in the ADS database. }
\label{z3950recfmt}
\begin{tabular*}{3.4in}{ll}

\hline
\noalign{\smallskip}

Value & Description

\\
\noalign{\smallskip}
\hline
\noalign{\smallskip}
B & Brief Records (Title, Authors)\\
F & Full Records (all available information)

\\
\noalign{\smallskip}
\hline
\end{tabular*}
\end{table}

\begin{table}
\caption[]{Record Syntax Supported in the ADS database. }
\label{z3950recsyn}
\begin{tabular*}{3.4in}{ll}

\hline
\noalign{\smallskip}

Value & Description

\\
\noalign{\smallskip}
\hline
\noalign{\smallskip}
1.2.840.10003.5.101 & SUTRS Records\\
1.2.840.10003.5.109.3 & HTML Records\\
1.2.840.10003.5.1000.147.1 & ADS Tagged Records

\\
\noalign{\smallskip}
\hline
\end{tabular*}
\end{table}

Each table notes which search fields support which attribute.  The
relationship attributes `equal' and `relevance' are used to search
without and with synonym replacement respectively.  If the structure
attribute `Phrase' is selected, the input is considered to be one
phrase and is not separated into individual words.  If the structure
attribute `Word' is selected and several words are specified, the
input is treated as if the structure attribute `Word List' were
specified.

As output, brief and full records are supported.  These record formats
are the same as in the short results list and in the full abstract
display as described above.  The supported record syntax is either
SUTRS (Simple Unstructured Text Record Syntax), HTML (HyperText Markup
Language), or the ADS tagged format.  In the HTML record syntax, links
to other supported ADS internal and external data sources are
included.

A description of this interface is available on-line at:

\begin{verbatim}
http://adsabs.harvard.edu/abs_doc/ads_server.html
\end{verbatim}

An example of the ADS Z39.50 interface can be accessed from the
Library of Congress Z39.50 Gateway at:

\begin{verbatim}
http://lcweb.loc.gov/z3950/
\end{verbatim}

\section{\label {cookies} Preferences}

The ADS user interface is customized through the use of so-called HTTP
persistent cookies (see \cite{cookies}).  These ``cookies'' are a
means of identifying individual users.  They are strings that are
created by the server software.  Web browsers that accept cookies
store these identification strings locally on the user's computer.
Anytime a user makes a request, the ADS software checks whether the
requesting browser is capable of accepting cookies.  If so, it sends a
unique string to the browser and asks it to store this string as an
identifier for that user.  From then on, every time the same user
accesses the ADS from that account, the browser sends this cookie back
to the ADS server.  The ADS software contains a database with a data
structure for each cookie that the ADS has issued.  The data structure
associated with each cookie contains information such as the type of
display the user prefers, whether tables should be used to format
data, which mirror sites the user prefers for certain external data
providers, the preferred printing format for ADS articles, and which
journal volumes the user has read.  It also can store the email
address of the user and a fax number, in case the user wants to
retrieve articles via email or fax.

The preference settings form includes a field for the user name as
well as the email address.  However, neither is necessary for the
functioning of any of the features of the ADS.  The system is
completely anonymous.  None of the information stored through this
cookie system is made available to anybody outside the ADS.  There is
no open interface to this database and the database files are not
accessible through the WWW.  Any particular user can only access their
own preferences, not the preferences set by any other user.

Most of these preferences can be set by the user through a WWW forms
interface.  Some fields in a user's preference record are for ADS
internal use only.  For instance the system is being used to display
announcements to users in a pop-up window.  The cookie database
remembers when the message was last displayed, so that each message is
displayed only once to each user.

This preference saving system also allows each user to store a query
form with filled-in fields.  This enables users to quickly query the
ADS for a frequently used set of criteria.

The cookie identification system is implemented as a Berkeley DBM
(DataBase Manager) database with the cookie as the record key.  The
data block that is stored in the database is a C structure.  The
binary settings (e.g. ``Use Tables'', or ``Use graphical ToC Page'')
are stored as bits in several preference bytes.  Other preferences are
stored as bytes, short integer, or long integer, depending on their
dynamic range.

HTTP cookies are based on host names.  Whenever a particular host is
accessed by the browser, the browser sends the cookie for that host to
the server software.  Since the ADS uses several host names as well as
mirror sites, we developed software that, on first contact between a
new user and the ADS, sets the same cookie for all our host names at
the main sites, as well as for the mirror hosts.  This is essential in
order to provide a seamless system with different servers handling
different tasks.

\section{\label {searchengine} Search Engine Details}

\subsection{General}

The search engine software is written in C.  It accepts as input a
structure that contains all the search fields and flags for the user
specified settings and filters.  For each search field that contains
user input a separate POSIX (Portable Operating System Interface)
thread is started that searches the database for the terms specified
in that field.  The results obtained for each term in that field are
combined in the thread according to the specified combination logic.
The resulting list of references is returned to the main thread.  The
main thread combines the results from the different field searches and
calculates the final score for each reference.  The final combined
list with the scores is returned to the user interface routines that
format the results according to the user specified output format.

\subsection{\label {searchalgorithm} Searching}

\subsubsection{Database Files}

The abstracts are indexed in separate fields: Author names, titles,
abstract text, and objects.  Each of these fields is indexed
similarly into an index file and a list file (see ARCHITECTURE).  The
index file contains a list of all terms in the field together with the
frequency of the term in the database, and the position and length of
two blocks of data in the list file.  One block contains all
references that include the exact word as specified.  The other block
contains all references that contain either the word or any of its
synonyms.

A search for a particular word in the index file is done through a
binary search in the index.  The indexes are resident in memory,
loaded during boot time (see section~\ref{optimization}).  Once the
word is found in the index, the position and length of the data block
is used to directly access the data block in the list file.  This data
block contains the identifier for each reference that contains the
search word.

If a quick update has occurred (see ARCHITECTURE) since the list file
was last built (indicated by a negative in part of the last
identifier), an auxiliary block of reference identifiers is read.  Its
position is contained in the structure of the last reference
identifier.  This auxiliary block is merged with the original one.

The identifier for each reference is the position of the bibliographic
code for that reference in the list of all bibliographic codes.  This
system saves one lookup of the identifier in a list of identifiers,
since the number can be used directly as an index in the array of
bibliographic codes.

\subsubsection{Synonym Searches}

As mentioned above, the index files contain information about two
blocks of data, the data for the individual word and for the synonym
group to which this word belongs.  When a search with synonym lookup
enabled is requested, the block of data for the whole synonym group is
used, otherwise the data for only the individual word is returned.
All the processing that enables these two types of searches is done
during indexing time, therefore the speeds for both searches are similar.

Even though our synonym list is quite extensive (see ARCHITECTURE) our
users will sometimes use words that are not in the database or the
synonym list.  In these cases the search software uses a stemming
algorithm from the Unix utility ispell to find the stem of the search
word and the searches for the word stem.  The indexing software has
indexed the stems of all words in the database together with their
original words (see ARCHITECTURE).  This word stemming is done as a
last resort if no regular match has been found in an attempt to find
any relevant references.

\subsubsection{\label {wildcard} Wildcard Searches}

In order to be able to search for families of words, a limited
wildcard capability is available.  Two wildcard characters are
defined:  The question mark `?' is used to specify a single wildcard
character and the asterisk `*' is used to specify zero or more wildcard
characters.  The `?' can be used anywhere in a word.  For instance
a search for {\it M1?} will find all Messier objects between M10 and
M19.  A search for {\it a?sorb} will find references with {\it absorb}
as well as {\it adsorb}.

The asterisk can only be used at the beginning or at the end of a
word.  For instance {\it 3C*} searches for all 3C objects.  {\it
*sorb} searches for words that end in {\it sorb} like absorb, desorb,
etc.  When synonym replacement is on, all their synonyms
(e.g. absorption) will be found as well.  The `?' and the `*' can be
combined in the same search string.

\subsection{Results Combining within a Field}

\subsubsection{Combining results}

As mentioned above, the user can select between four types of
combination methods: ``OR'', ``AND'', simple logic, and full boolean
logic.  For the first three cases, the references for all search terms
are retrieved and sorted first.  The reference lists
are then merged by going through the sorted reference lists
sequentially and synchronously and selecting references according to
the chosen logic.

The search algorithm for the full boolean logic is different.  The
boolean query is parsed from left to right.  For each search term a
function is called that finds the references for this term.  A search
term is either a search word, a phrase, or an expression enclosed in
parentheses.  If the search term contains other terms (if it is
enclosed in parentheses), the parsing function is called recursively.

The next step is to determine the boolean operator that follows the
search term, and then to evaluate the next search term after the
operator.  Once the reference lists for the two terms and the
combining operator are determined, the two lists are combined
according to the operator.  This new list is then used as the first
term of the next expression.

If the boolean operator is `OR', the combining of the lists is
deferred, and the next operator and search term are evaluated.  This
ensures the correct precedence of `OR' and `AND' operators.

The `NOT' operator is implemented by getting the reference list for
the term, making a copy of all references in the database, and then
removing the references from the search term from the complete list.
This yields a very large list of references.  If the first search term
in a boolean expression is a `NOT' term, the search will take very
long, because this large list has to be propagated through all the
subsequent parsing of the boolean expression.  Care should therefore
be taken to put a `NOT' term to the right of at least one other term,
since processing is done left to right.

As an example, figure~\ref{pseudocode} shows the processing of the
boolean expression mentioned in section~\ref{dataretrieval}:

\begin{figure}
\resizebox{\hsize}{!}{\includegraphics{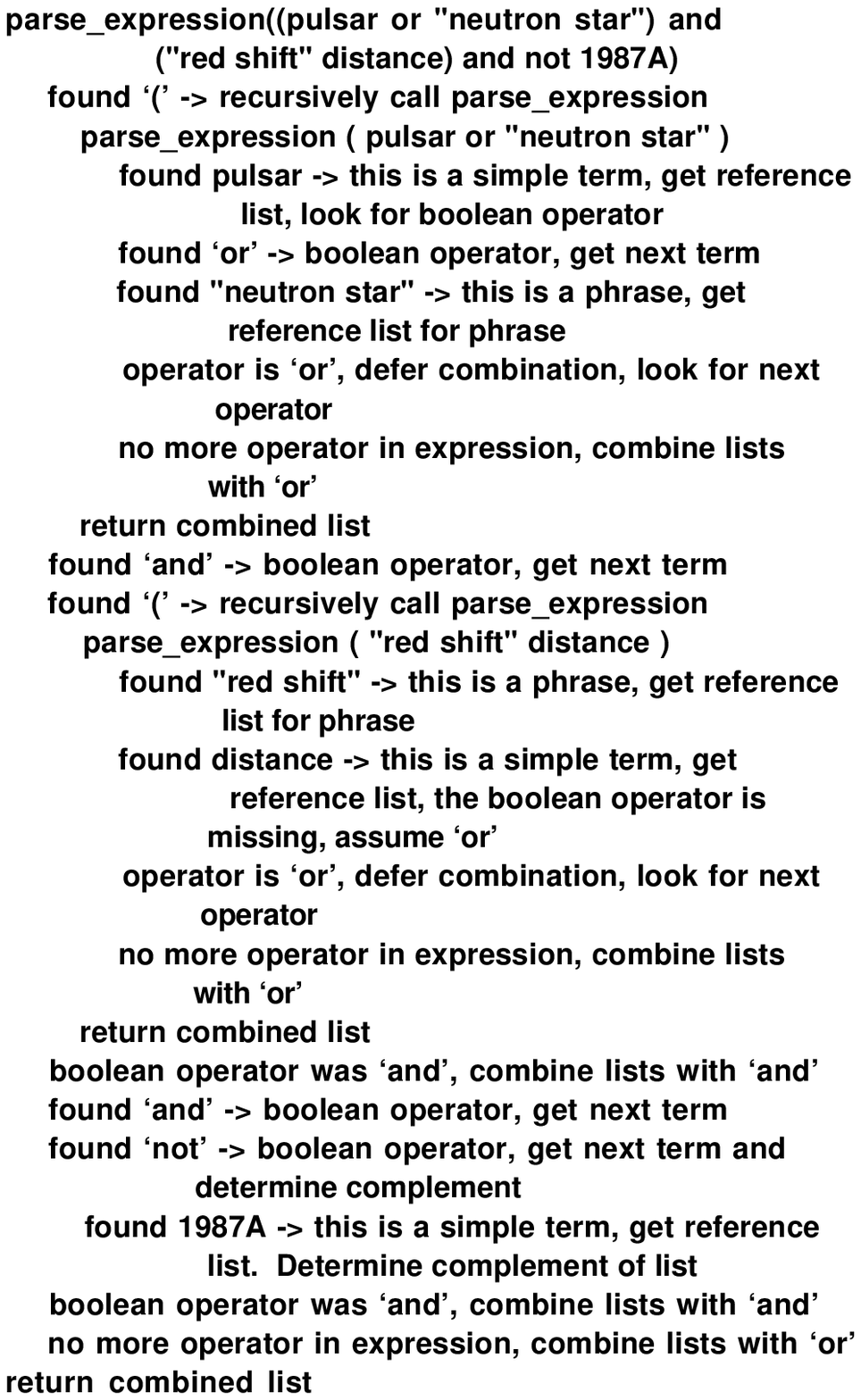}}
\caption[]{Pseudo code for parsing full boolean search  expression. }
\label{pseudocode}
\end{figure}

\begin{verbatim}
(pulsar or ``neutron star'') and
   (``red shift'' distance) and
   not 1987A
\end{verbatim}

\subsubsection{Scoring}

In addition to the information about the references for each word, the
index file also contains its frequency in the database.  The frequency
is already pre-calculated as int(10000/(log(frequency))) during
indexing (see ARCHITECTURE).  This saves considerable time during
execution of the search engine since all server calculations can be
done as integer computations, no floating point operations are
necessary.  During the first part of the search, this frequency is
attached to each retrieved reference.  In the next step, the retrieved
references are combined according to the selected combination logic
for that field.

For `OR' combination logic, the lists retrieved for each word are
merged and uniqued.  As described in section~\ref{optimization}, this is
done by going through the sorted reference lists synchronously and
adding each new reference to the output list.  The score for that
reference is determined by adding up the frequencies from each of the
lists for weighted scoring, or by setting a score equal to the number
of matched words for proportional scoring.

For `AND' combination logic, only references that appear in every one
of the lists  are selected.  Each of these references receives a score
of 1.

For simple logic and full boolean logic, the score for the returned
references is determined only from the search terms that were combined
with `OR'.  All words that have mandatory selection criteria (prefixed
by `+' or `--' in simple logic, and combined with `AND' or `NOT' in
full boolean logic) do not affect the final score.

\subsection{Combining Results among Fields}

\subsubsection{Combining}

After the POSIX threads for each search field are started, the main program
waits for all threads to complete the search.  When all searches are
completed the search engine combines the results of the different
searches according to the selected settings.  If for instance one
field was selected as required for selection in the settings section
of the query form, only references that were found in the search for
that field will be in the final reference list.  The combined list is
then uniqued and sorted by score.  The resulting list of
references is passed back to the user interface software.

If the user did not specify any search terms, a date range has to be
selected.  The software queries the database for all references in
the selected date range and uses this list for further processing,
e.g. filtering (see section~\ref{filters}).

\subsubsection{Scoring}

The score for each reference in the final results list is determined
by adding the scores from each list multiplied by the user specified
weight for each field and then normalizing the score such that a
reference that matches all search terms from all fields receives a
score of 1.

\subsection{\label{filters} Selection Filters}

After the search is completed according to the specified search words
and the settings that control the combination logic and the scoring
algorithms, the resulting list of references can be filtered according
to several criteria.  During the design of the software a decision had
to be made whether to filter the results while selecting the
references or after completing the search.  The first approach has the
advantage that the combining of the selected references will be faster
because fewer references need to be combined.  The second approach has
the advantage that the first selection is faster.  We chose the second
approach because, except for selecting by publication date, only a
small number of queries use filtering (see section~\ref{accessstats}).
Because of that, filtering by publication date is done during
selection of the references, while the other filtering is done after
the search is completed.

References can be filtered by five criteria:
\begin{verbatim}
1.  Entry date in the data base
2.  Minimum score
3.  Journal
4.  Available links
5.  Group membership
\end{verbatim}

1. + 2.  Entry date and minimum score.  These two filters can be used
to query the database automatically on a regular basis for new
information that is relevant to a selected topic.  The user can build
a query form that returns relevant references, and then save this
query form locally.  This query form can then be sent to the ADS email
interface (see above) on a weekly or monthly basis.  By specifying an
entry day of -7 for instance, the query will retrieve all references
that fit the query and that were entered within the last seven days.
The minimum score can be used to limit the returned number of
references to only the ones that are really relevant.  The references
are returned via email as described in section~\ref{email} about the email
interface.

3.  Journal filter.  This filter allows the user to select references
from individual journals or groups of journals.  Available options
for this filter are:
\begin{verbatim}
a.  All journals (default)
b.  Refereed journals only
c.  Non-refereed journals only
d.  Selected journals
\end{verbatim}

If the last option is selected, the user can specify one or more
journal abbreviations (e.g. ApJ, AJ (Astronomical Journal)) that
should be selected.  More than one abbreviation can be specified by
separating them with semicolons or blanks.  The filter for journals
can also include the volume number (but not the page number).  The
journal abbreviation is compared with the bibliographic codes over the
length of the specified abbreviation.  For instance if the user
specifies {\it ApJ}, the system selects all articles published in the
ApJ, ApJ Supplement and ApJ Letters.  {\it ApJ..} will select only
articles from the ApJ and the ApJ Letters.  A special abbreviation,
{\it ApJL} will select only articles from the ApJ Letters.  If a
journal abbreviation is specified with a prepended `--', all
references that are NOT from that journal are returned.  The journal
abbreviations (or bibstems) used in the ADS are available at:

\begin{verbatim}
http://adsdoc.harvard.edu/abs_doc/
       journal_abbr.html
\end{verbatim}

4.  Available links.  This filter allows the user to select references
that have specific other information available.  The returned
references can be filtered for instance to include only references
that have data links or scanned articles available.  As an example,
a user needs to find on-line data about a particular object.  A search
for that object in the object field and a filter for references with
on-line data returns all articles about that object that link to
on-line data.

5.  Groups.  We provide the capability to build a reference collection
for a specific subset of references.  This can be either articles
written by members of a particular research institute or about a
particular subject.  Currently there are 5 groups in the ADS.  We
encourage larger institutes or groups to compile a list of their
references and send it to us to be included in the list of groups.

\section{\label{optimization} Optimizations}

The search engine is entirely custom-built software.  As mentioned in
the introduction, the first version of the Abstract Service used
commercial database software.  Because of too many restrictions and
serious performance problems, a custom-designed system was
developed.  The main design goal was to make the search engine as fast
as possible.  The most important feature that helped speed up the
system was the use of permanent shared memory segments for the search
index tables.  In order to make searching fast, these index tables
need to be in Random Access Memory.  Since they are tens of megabytes
long, they cannot be loaded for each search.  The use of permanent
shared memory segments allows the system to have all the index tables
in memory all the time.  They are loaded during system boot.  When a
search engine is started, it attaches to the shared segments and has
the data available immediately without any loading delays.  The shared
segments are attached as read-only, so even if the search engine has
serious bugs, it cannot compromise the integrity of the shared
segments.  Using shared segments with the custom-built software
improved the speed of a search by a factor of 2 -- 20, depending on the
type of search.

Access to the list files (see section~\ref{searchalgorithm}) was
optimized too.  These files cannot be loaded into memory since they
are too large (each is over one hundred megabytes in size).  To
optimize access to these files, they are memory mapped when they are
accessed for the first time.  From then on they can be accessed as if
they were arrays in memory.  The data blocks specified in the index
tables can be accessed directly.  Access is still from file, but it is
handled through the paging in the operating system, rather than
through the regular I/O system, which is much more efficient.

Once the search engine was completed and worked as designed, it was
further optimized by profiling the complete search engine and then
optimizing the modules that used significant amounts of time.  Further
analysis of the performance of the search engine revealed instances
where operations were done for each search that could be done during
indexing of the data and during loading of the shared segments.
Overall these optimizations resulted in speed improvements of a factor
of more than 10 over the performance of the first custom-built
version.  These optimizations were crucial for the acceptance of the
ADS search system by the users.

In order to further speed up the execution, the search engine uses
POSIX threads to exploit the inherent parallel nature of the search
task.  The search for each field, and in the case of the object field
for each database, is handled by a separate POSIX thread.  These
threads execute in parallel, which can provide speedups in our
multiprocessor server.  Even for single processor systems this will
provide a decrease in search time, since each thread sometimes during
its execution needs to wait for I/O to complete.  During these times
other threads can execute and therefore decrease the overall execution
time of a search.

Another important part of the optimization was the decision on how to
structure the index and list files.  The index files contain the
word frequency information that is used to calculate scores for the
weighted scoring (see section~\ref{abstracts}c).  The score for a matching
reference is calculated from the inverse logarithm of the frequency of
the word in the database.  This requires time consuming floating point
calculations.  To avoid these calculations during the searches, the
floating point arithmetic is done at indexing time.  The index file
contains the inverse log of the word frequency multiplied by a
normalization factor of 10,000.  This allows all subsequent
calculations to be done in integer arithmetic, which is considerably
faster than floating point calculations.

Another optimization was to pre-compile the translation rules (see
section~\ref{abstracts}).  These translation rules are pre-compiled and
stored in a shared memory segment to which the search process
attaches.  This allows for faster execution of these pattern matching
routines.

Overall, these optimizations improved the speed of the searches by two
orders of magnitude between the original design using a commercial
database and the current software.

\section{\label {accessstats} Access Statistics}

The ADS software keeps extensive logs about the use of the search
and access software.  In this section, usage statistics for the search
software and for access patterns to the Article Service are reported.
If not otherwise indicated, the statistics in this section are for the
one-year period from 1 April 1998 through 31 Mar 1999.

\subsection{Abstract Service}

The ADS is accessed by users from many different countries.  In the
one-year period of this section the ADS was accessed by 127,000
different users, using 100,000 different hosts from 112 different
countries.  An individual user is defined as having a unique cookie
(see section~\ref{cookies}).  Users without cookies are distinguished
by the hostnames from which the requests came.  This may overestimate
the number of users, since some users may have more than one cookie,
for instance when accessing the ADS from home.  The number of
different hosts is a lower limit of the number of users.  Many hosts
are used by multiple users, so the real number is certainly
considerably higher than that.  The development of the number of users
and queries over the life of the ADS is described in OVERVIEW.  This
section describes some more detailed investigations of the access
statistics.

The total number of users at first comes as quite a surprise.  The
number of working astronomers in the world is probably between 10,000
and 20,000.  The number of ADS users is much larger than that.  This
is probably due to several factors.  First, there are certainly many
accidental users.  They somehow find our search page through some
link, execute a query to see what they get back, and then never come
back because it is of no interest to them.

Other possible users are media people.  There are certainly many
reporters occasionally looking up something in astronomy.  I have
spoken with several of them that use the ADS occasionally for that.

Another group of users are amateur astronomers.  The ADS was described
in Sky \& Telescope by \cite{1996S&T....92...81E} a few years ago.
this has certainly made amateur astronomers aware of this resource.
The number of amateur astronomers world-wide is certainly in the
millions, so they comprise a potentially large number of users.

Another large group of users visits the ADS through links from other
web sites.  One particularly popular one is NASA's Image of the Day,
which frequently includes links to abstracts or articles in the ADS.
Since this NASA page is visited by millions of people, a large number
of them will access the ADS through these links.

\begin{figure}
\resizebox{\hsize}{!}{\includegraphics{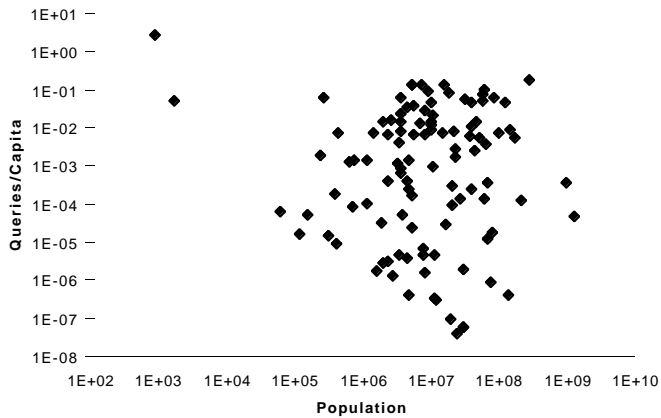}}
\caption[]{Number of queries per person per year in each country as a function of the population for the country. }
\label{qupop}
\end{figure}

The use of the ADS in different countries depends on several
factors.  One of these is certainly the population of the country.
Figure~\ref{qupop} shows the number of queries per capita as a
function of the population of the country (\cite{ciafacts}).  There
seems to be an upper limit of about 0.1 -- 0.2 queries per person per
year.  The one exception is the Vatican with almost 3 queries per
person per year.  This is understandable since the Vatican has an
active Astronomy program, which generates a large number of queries
for a small population.

\begin{figure}
\resizebox{\hsize}{!}{\includegraphics{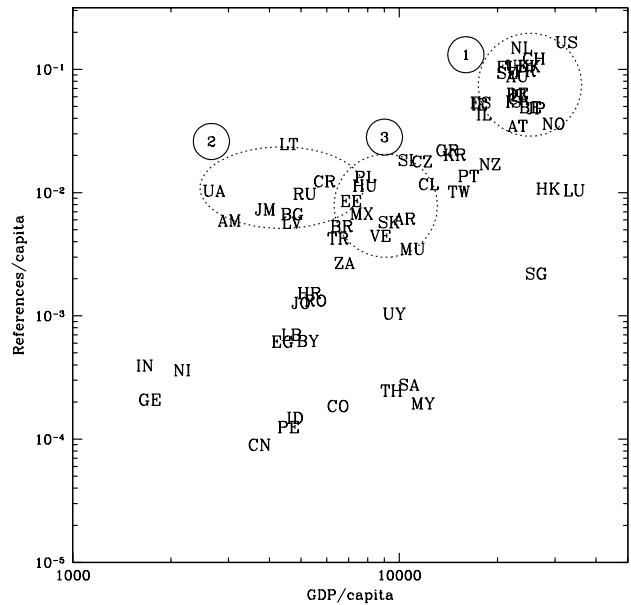}}
\caption[]{Number of references retrieved per person as a function of the Gross Domestic Product (GDP) per person for each country. }
\label{gdprefs}
\end{figure}

Another factor for querying the ADS is the funding available for
Astronomy in a country, and the available infrastructure to do
astronomical research.  Figure~\ref{gdprefs} shows the number of
references retrieved per capita as a function of the Gross Domestic
Product (GDP, \cite{ciafacts}) of the country.  The symbols are the Internet
codes for each country.

\begin{figure}
\resizebox{\hsize}{!}{\includegraphics{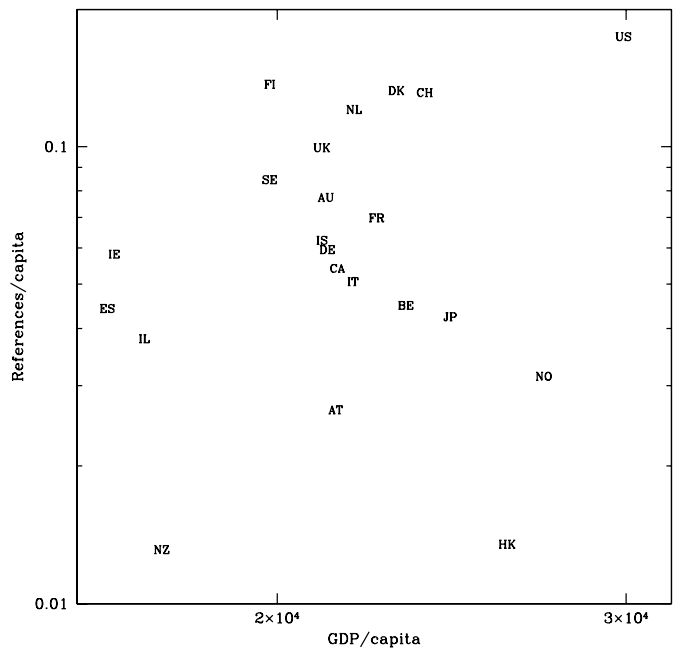}}
\caption[]{Number of references retrieved per person as a function of
the Gross Domestic Product (GDP) per person.  This figure is a closeup
of area 1 in figure~\ref{gdprefs}. }
\label{gdprefsclose}
\end{figure}

The highly industrialized countries cluster in the upper right part of
the plot (area 1).  A closeup of this region is shown in
figure~\ref{gdprefsclose}.  Other clusters are the countries of the
former Soviet Union (area 2), and Central and South American countries
(area 3).  The high number of references retrieved per capita combined
with the lower GDP per capita of the former Soviet Union is probably
due to a recent decline in GDP, but a still existing infrastructure
for astronomical research.

\begin{figure}
\resizebox{\hsize}{!}{\includegraphics{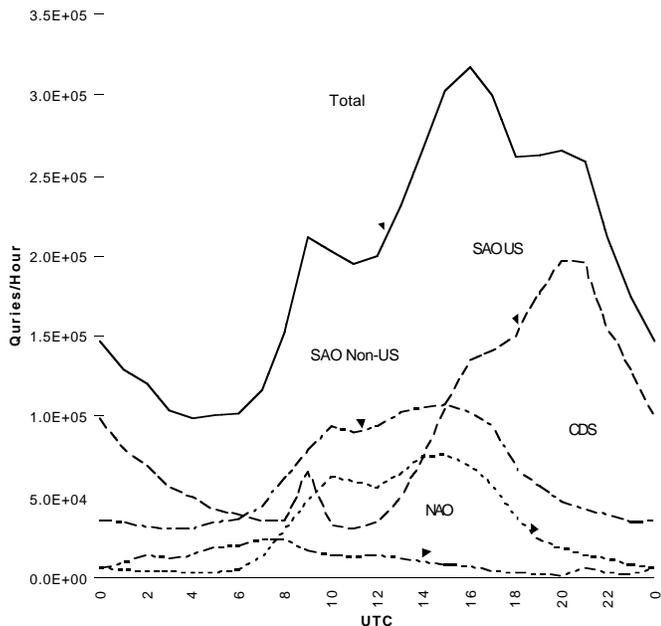}}
\caption[]{Number of queries per hour as a function of the time of day
for the main SAO site and the mirror sites in France and Japan. }
\label{24hqu}
\end{figure}

The ADS is used 24 hours per day.  The distribution of queries
throughout the day is shown in figure~\ref{24hqu}.  This figure shows
the number of queries at the two largest mirror sites, as well as the
queries at the main ADS site.  The usage distribution data are for the
time period from 1 November 1998 to 31 March 1999, not the full year,
to avoid complications due to different periods where daylight savings
time is in effect.  The queries at the main site are separated into
US users and non-US users on the basis of their Internet hostnames.
All the individual curves show a distinct two-peaked basic shape, with
additional smaller peaks in some cases.  This distribution of queries
over the day shows the usage throughout a workday, with a small
minimum during lunch hour.  The SAO-US distribution does not show a
real minimum between the two peaks, presumably because of the
distribution of US researchers over three time zones.

There are three features in this figure that deserve special notice.
The first is that the shape of the accesses to the ADS mirror in
France is the same as the shape of the non-US access to the SAO site.
This indicates that the large majority of the non-US use on the SAO
site is from European users.

This non-US usage is about 50\% higher than the total usage of
the ADS mirror site at the CDS in France.  The reason for this is most
probably the fact that the connectivity within Europe is not yet very
good.  We know that for instance that our users in England and Sweden
have better access to the main ADS site in the USA than to our mirror
site in France.  The same is true for other parts of Europe.

Another reason for the use of the USA site by European users is the
fact that our European mirror sites do not yet have the complete set
of scanned articles on-line.  This forces some users to access the
main ADS site in order to retrieve scanned articles.

Second, there is a slight peak in the distribution of queries to the
NAO mirror in Japan around 21:00 UTC (Universal Time Coordinated,
formerly Greenwich Mean Time).  This is probably due to US west
coast users using the Japanese mirror site instead of the US site.
The access to Japan is frequently very fast and response times from
Japan may be better than from SAO during peak traffic times.

Third, there is a distinct peak in the SAO-US usage at 9:00 UTC.  This
feature was so unusual that we tracked down the reason for it.  It
turns out that one of our users has set up web pages that include
about 200 links to ADS abstracts.  He had set up a link verifier that
every night at 9:00 UTC checked all the links on his pages.  This meant
that the link verifier executed 200 queries every night at the same
time, which showed up in this evaluation of our access statistics.

\begin{figure}
\resizebox{\hsize}{!}{\includegraphics{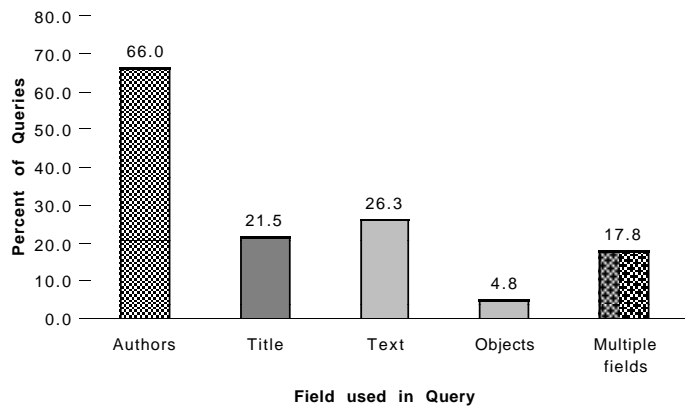}}
\caption[]{A histogram of the relative usage of the different search fields (authors, objects, title, text) and the use of multiple fields. }
\label{fieldhist}
\end{figure}

The following section shows statistics of how our users use the
different capabilities of the ADS query system.
Figure~\ref{fieldhist} shows a histogram of the relative usage of the
different search fields (authors, objects, title, text).  It shows
clearly that the majority of queries are queries by author name
(66\%).  Object names are used in fewer than 5\% of the queries.  The
title field is used in about 21\% of the queries, and the text field
in 26\%.  Queries that use more than one field make up about 18\% of
the total.  This usage pattern justifies for instance including tables
of contents (ToCs) in the database that do not have abstracts for searching.
Since a large part of the usage is through author and title queries,
such ToC entries will still be found.

\begin{figure}
\resizebox{\hsize}{!}{\includegraphics{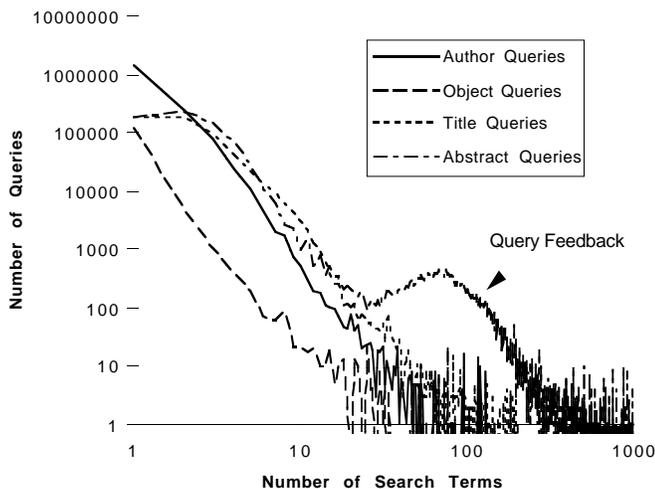}}
\caption[]{The number of queries in the period of 1 April 1998 to 31 March 1999 as a function of the number of query items in each input field. }
\label{numsearchitems}
\end{figure}

Figure~\ref{numsearchitems} shows the number of queries as a function
of the number of query items in each input field.  The query frequency
generally decreases exponentially with increasing number of search
terms.  For title and text queries, the frequency is approximately
constant up to 3 query words, before the frequency starts to decrease.
For abstract queries there is a significant increase in frequency of
queries with more than 20 query words, for title queries there is a
similar increase for queries with more than about 8 query words.  This
is due to queries generated through the query feedback mechanism which
allows the user to use a given abstract and its title as a new
query.

\begin{figure}
\resizebox{\hsize}{!}{\includegraphics{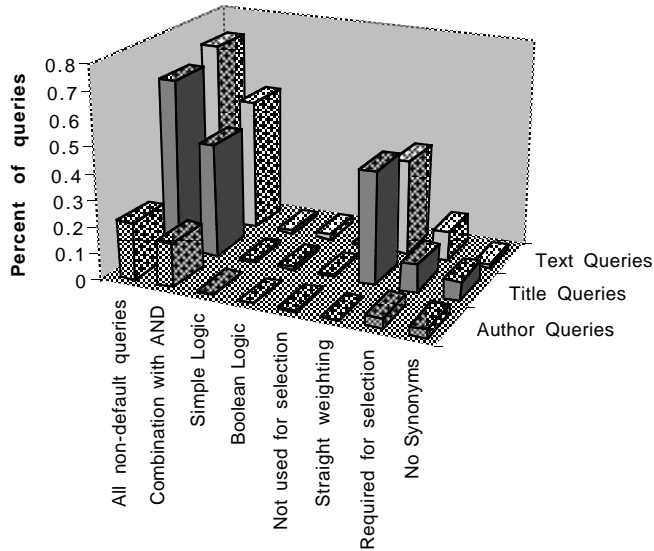}}
\caption[]{Percentage of non-default settings for the different available settings and query fields. }
\label{settingsuse1}
\end{figure}

\begin{figure}
\resizebox{\hsize}{!}{\includegraphics{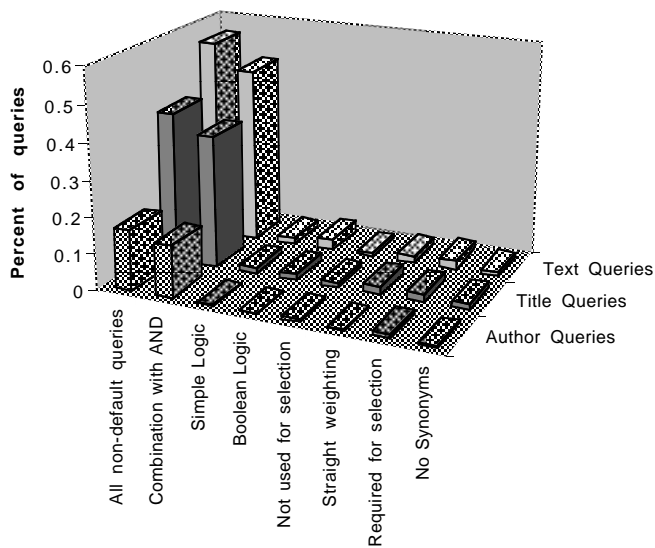}}
\caption[]{Percentage of non-default settings for the different available settings and query fields.  This plot excludes the queries from NASA Techreports and SPIE. }
\label{settingsuse2}
\end{figure}

Figures~\ref{settingsuse1} and \ref{settingsuse2} show the usage of
non-default query settings (see section~\ref{searchalgorithm}).  The
default settings were chosen to suffice for most queries.
Figure~\ref{settingsuse1} shows the percentage of non-default settings
for the different settings and query fields available.  It shows that
29\% of author queries, 78\% of title queries, and 85\% of text
queries use non-default settings.  This was at first disappointing,
because it suggested that the default settings might not be a
reasonable selection of settings.  The two main settings that were
non-default were combining words with ``AND'' (see
section~\ref{abstracts}.b.ii), and disabled weighted scoring (see
section~\ref{abstracts}.c).  On closer examination of the statistics it
turns out that the straight weighting settings come from mainly two
systems, the NASA Techreports and the International Society for
Optical Engineering (SPIE).

Both of these systems use our Perl scripts (see
section~\ref{perlaccess}) to access the ADS database.  They do not set
our normal default settings during these queries.
Figure~\ref{settingsuse2} shows the non-default settings for all
queries that did not come from either of these two servers.  There is
still a small percentage of queries that use straight weighting,
probably mostly due to other systems that use our Perl script
interface routines.

The one remaining non-default setting that is used frequently is the
combination of words with ``AND''.  We believe that the ``OR''
combination as default is more useful since it returns more
information.  The beginning of the list of returned references is the
same, regardless of whether ``AND'' or ``OR'' combination is selected,
since references that match all words are sorted to the beginning of
the list.  When ``OR'' combination is selected, partial matches will
be returned after the ones with perfect matches.  This is desirable
since there may be relevant references that for some reason do not
match all query words.

The other selecting mechanism that is available is the filtering of
references according to what other information is available for a
reference.  The usage of the filtering is shown in
table~\ref{filteruse}.  About 10\% of the total queries use the filter
option.  Almost all of these filter by journal or select refereed
journals only.  The sum of the numbers for required data types adds up
to more than the number for ``Required data'', since more than one
data type can be selected.

\begin{table}
\caption[]{Filter requests during the period of 1 April 1998
to 31 March 1999. }
\label{filteruse}
\begin{tabular*}{3.4in}{llr}

\hline
\noalign{\smallskip}

Filter Type & Required Data Type & Queries

\\
\noalign{\smallskip}
\hline
\noalign{\smallskip}
Total queries & ~ & 2754405\\
Non-standard queries & ~ & 286341\\
Selected journal &~  & 158581\\
Refereed journals & ~ & 96270\\  
Non-refereed journals & ~ & 1616\\
Data available & ~ & 6381\\
Required data & ~ & \\
~ & Printable Articles & 2921\\
~ & Scanned Articles & 1951\\
~ & Electronic Articles & 1690\\
~ & Abstracts & 1382\\
~ & Planetary Data System & 834\\
~ & Planetary Nebulae & 667\\
~ & Citations & 615\\
~ & Table of Contents & 506\\
~ & References & 459\\
~ & Author Comments & 393\\
~ & On-line Data & 360\\
~ & SIMBAD Objects & 269\\
~ & NED Objects & 212\\
~ & Library Entries & 204\\
~ & Mail Order & 201\\
~ & Associated Articles & 83

\\
\noalign{\smallskip}
\hline
\end{tabular*}
\end{table}

Table~\ref{datalinkuse} shows the number of links available and the
usage pattern of the data links that the ADS provides.  The highest
usage is access to the abstracts, followed by the links to full text articles,
links to citations, and links to on-line electronic
articles.  Reference links and links to SIMBAD objects are next.

\begin{table}
\caption[]{Link types and their accesses during the period of 1 April 1998
to 31 March 1999.  Numbers of links available are as of July 1999.
}
\label{datalinkuse}
\begin{tabular*}{3.4in}{lrr}

\hline
\noalign{\smallskip}

Links & Nr. Links & Nr. Accesses

\\
\noalign{\smallskip}
\hline
\noalign{\smallskip}
Abstracts & 941,293 & 1,608,726\\
Scanned Articles & 138,785 & 526,872\\
Printable Articles & 40,928 & 254,881\\
(Postscript and PDF)\\
Electronic Articles & 125,933 & 186,067\\
Citations & 195,192 & 77,316\\
References & 135,474 & 36,969\\
SIMBAD Objects & 110,308 & 23,505\\
On-line Data & 5,728 & 9,799\\
NED Objects & 31,801 & 6,144\\
Mail Order & 247,282 & 3,520\\
Library Entries & 18,746 & 1,645\\
Tables of Contents & 5,792 & 1,233\\
Author Comments & 203 & 313\\
Associated Articles & 2765 & 169\\
Planetary Nebulae Data & 281 & 143\\
\noalign{\smallskip}
\hline
\end{tabular*}
\end{table}

\subsection{Article Access Statistics}

The ADS Article Service provides access to full journal articles.  The
usage statistics should show how astronomy researchers read and use
journal articles.  In this section we describe a few of the statistics
of the article server.  More statistics on the usage of the scanned
articles are described in OVERVIEW.

\begin{figure}
\resizebox{\hsize}{!}{\includegraphics{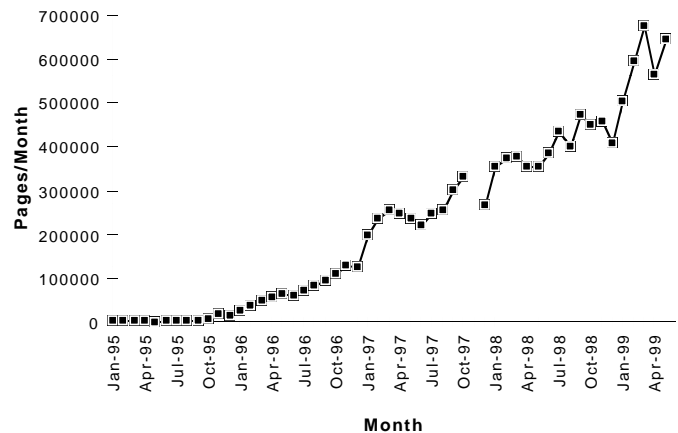}}
\caption[]{Number of pages of scanned articles retrieved through the life of the ADS Article Service. }
\label{nrarticlepages}
\end{figure}

\begin{figure}
\resizebox{\hsize}{!}{\includegraphics{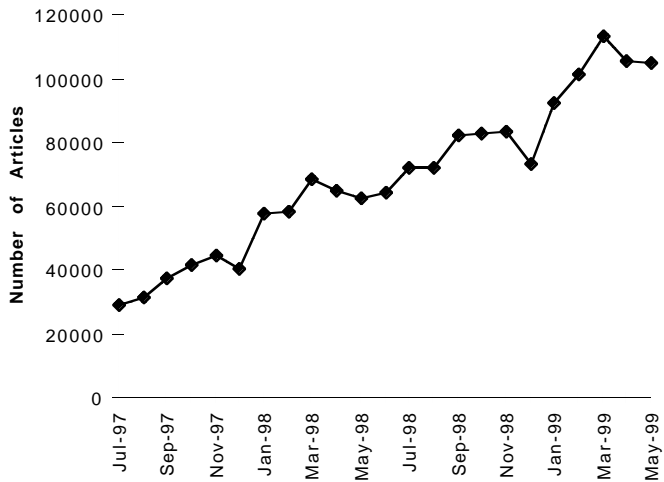}}
\caption[]{Number of full text articles retrieved by ADS users.  These numbers include the scanned articles at the ADS, as well as articles at the sites of the different journals that were requested through ADS links. }
\label{nrarticles}
\end{figure}

Figure~\ref{nrarticlepages} shows the number of pages of scanned
articles retrieved over the life of the ADS, figure~\ref{nrarticles}
shows the number of articles retrieved.  The number of articles
represents the sum of the selected links to on-line electronic
articles, PDF and Postscript articles at the journals, and scanned
articles at the ADS.

Both the number of pages and the number of articles retrieved is
steadily increasing.  This is due to both the increased coverage in
the ADS of scanned journals and the increase in the number of users
that use the system.

Table~\ref{atclretrieval} shows the number of retrievals in the
various formats.  Postscript is a printer control language developed
by Adobe (see \cite{postscript}).  Postscript Level 1 is the first
version of the Postscript language.  It generates much larger files
than Level 2 Postscript.  Some older printers can process only Level 1
Postscript files.  PDF (Portable Document Format) is a newer page
description format, also developed by Adobe.  PCL (Printer
Control Language) is a printer control language developed by Hewlett
Packard.  It is used in low end PC printers.  Low
resolution is 200 dpi for Postscript and PDF, and 150 dpi for PCL.
High resolution is 600 dpi for Postscript and PDF, 300 for PCL.

\begin{table}
\caption[]{Article retrieval by format type for March 1999, March 1998,
and March 1997.  PDF format and GIF Thumbnails were not yet available
in March 1997. }
\label{atclretrieval}
\begin{tabular*}{3.4in}{p{0.35\linewidth}rrr}

\hline
\noalign{\smallskip}

Article Type & \multicolumn{3}{c}{Number of Retrievals}\\
& March 99 & March 98 & March 97

\\
\noalign{\smallskip}
\hline
\noalign{\smallskip}
Postscript Level 1 & 476 & 557 & 644\\
(Low Resolution)\\
Postscript Level 2 & 25,664 & 13,031 & 11,189\\
(Low Resolution)\\
Postscript Level 2 & 10,472 & 8,291 & 6,435\\
(High Resolution)\\
PDF & 3,266 & 620 & n/a\\
(Low Resolution)\\
PDF & 7,049 & 1008 & n/a\\
(High Resolution)\\
PCL & 14 & 73 & 72\\
(Low Resolution)\\
PCL & 53 & 111 & 132\\
(High Resolution)\\
GIF Thumbnails & 13,777 & 7,378 & n/a\\

\\
\noalign{\smallskip}
\hline
\end{tabular*}
\end{table}

The majority of retrievals are of medium resolution
Postscript files.  This is the default setting in the ADS Article
Service.  The number of Postscript Level 1 articles (compatible with
older printers, but much larger file sizes) retrieved is low compared
with Level 2 Postscript articles, and slowly declining.  The number of
PCL articles retrieved is even smaller and also declining.  The number
of PDF articles retrieved was slowly increasing throughout 1998.  It
has increased much more rapidly in 1999.  In early 1998 less than 15\%
of the high resolution articles were retrieved as PDF files.  This
fraction increased to 40\% by March, 1999.

\section{\label {future} Future Plans}

The ADS Abstract Service is only seven years old, but it is already an
indispensable part of the astronomical research community.  We get
regular feedback from our users and we implement any reasonable
suggestions.  In this section we mention some of the plans for
improvements that we are currently working on to provide even more
functionality to our users.

\subsection{Historical Literature}

One important part of the ADS Digital Library will be the historical
literature from the 19th century and earlier.  This part of the early
literature is especially suited for being on-line in a central digital
library.  It is not available in many libraries, and if available is
often in dangerously deteriorating condition.  The access statistics
of the ADS show that even old journal articles are accessed regularly
(see OVERVIEW).  The ADS is working on scanning this historical
literature through two approaches: Scanning the journals, and scanning
the observatory literature.

1.  We are in the process of scanning the historical journal
literature.  We already have most of the larger journals scanned
completely.  Table~\ref{scannedjour} shows how much we have scanned of
each of the journals and conference proceedings series for which we
have permission.  The oldest journal we have scanned completely is the
Astronomical Journal (Vol. 1, 1849).  We plan to have the Monthly
Notices of the Royal Astronomical Society on-line completely by early
in 2000.  After that we plan to scan the oldest astronomical journal,
Astronomische Nachrichten (Vol. 1, 1821), Icarus, Solar Physics,
Zeitschrift f\"ur Astrophysik, Bulletin of the American Astronomical
Society, and L'Observateur, and the conference series of IAU Symposia.
Other journals that we plan to scan are the other precursor journals
for Astronomy \& Astrophysics, if we can obtain permission to do so.
We will also scan individual conference proceedings for which we can
obtain permission.

\begin{table*}
\caption[]{Scanned journals in the ADS database. }
\label{scannedjour}
\begin{tabular*}{7.0in}{p{0.5\linewidth}ll}

\hline
\noalign{\smallskip}

Journal & Scanned Volumes & Publication Dates

\\
\noalign{\smallskip}
\hline
\noalign{\smallskip}
Acta Astronomica & 42-48 & 1992-1998\\
Annual Reviews of Astronomy and Astrophysics & 1-33 & 1962-1995\\
Annual Reviews of Earth and Planetary Sciences & 1-23 & 1973-1995\\
Astronomical Journal & 1-114 & 1849-1997\\
Astronomical Society of the Pacific Conference Series & 1-5, 7-22, 24-63, 65-69& 1988-1994\\
Astronomy and Astrophysics & 1-316 & 1969-1996\\
Astronomy and Astrophysics Supplement Series & 1-120 & 1969-1996\\
Astrophysical Journal & 1-473 & 1895-1996\\
Astrophysical Journal Letters & 148-473 & 1967-1996\\
Astrophysical Journal Supplement Series & 1-107 & 1954-12/1996\\
Baltic Astronomy & 1-5 & 1992-1996\\
Bulletin Astronomique de Belgrade & 153-155 & 1996-1997\\
Bulletin of the Astronomical Institute of Czechoslovakia & 5-6, 9-42 & 1954-1955, 1958-1991\\
Bulletin of the Astronomical Society of India & 8-23 & 1980-1995\\
Journal of the Korean Astronomical Society & 1-29 & 1968-1996\\
Journal of the British Astronomical Association & 92-107 & 1981-1997\\
Meteoritics and Planetary Science & 1-33 & 1953-1998\\
Monthly Notices of the Royal Astronomical Society & 1, 100-301 & 1827, 1950-12/1998\\
AAS Photo Bulletin & 1-43 & 1969-1986\\
Publications of the Astronomical Society of Australia & 1, 3-13 & 1967, 1976-1996\\
Publications of the Astronomical Society of Japan & 1-50 & 1949-1998\\
Publications of the Astronomical Society of the Pacific & 1-109 & 1889-1997\\
Reviews in Modern Astronomy & 1-9 & 1988-1996\\
Revista Mexicana de Astronomia y Astrofisica & 1-32 & 1974-1996\\
Revista Mexicana de Astronomia y Astrofisica Ser. de Conf. & 1-6 & 1995-1997\\
Skalnate Pleso, Contributions & 11-14, 16-19, 21-28 & 1983-1986, 1987-1990, 1991-1998\\
\noalign{\smallskip}
\hline
\end{tabular*}
\end{table*}

2.  One very important part of the astronomical literature in the 19th
century and earlier were the observatory publications.  Much of the
astronomical research before the 20th century was published in such
reports.  We are currently collaborating with the Harvard library in a
project to make this part of the astronomical literature available
through the ADS.  The Harvard library has a grant from the National
Endowment of the Humanities to make preservation microfilms of this
(and other) literature.  This project is generating an extra copy of
each microfilm.  We will scan these microfilms and produce electronic
images of all the microfilmed volumes.  The resolution of the
microfilm and the scanning process is approximately equivalent to our
600 dpi scans.  This project, once completed, will provide access to
all astronomical observatory literature that is available at the
Harvard library and the library of the Smithsonian Astrophysical
Observatory from the 19th century back.  For the more recent
observatory literature we will have to resolve copyright issues before
we can put it on-line.

In order to complete our data holdings, we still need issues of
several journals for scanning.  A list of journals and volumes that we
need is on-line at:

\begin{verbatim}
http://adsabs.harvard.edu/pubs/
       missing_journals.html
\end{verbatim}

In order to feed the pages through a sheet feeder, we need to cut the
back of the volumes to be scanned.  If they have not been bound
before, they can be bound after the scanning.  If they have been bound
before, there is not enough margin left to bind them again.  If you
have any of the journals/volumes that we need, and you are willing to
donate them to us, please contact the first author of this article.
We would like to have even single volumes of any of the missing journals.

\subsection{Search Capabilities}

The capabilities of the search system and user interface have been
developed in close cooperation with our users.  We always welcome
suggestions for improvements and usually implement reasonable
suggestions very quickly (within days or a few weeks).  Because of
this rapid implementation of new features we have hardly any backlog
of improvements that we want to implement.  There are currently two
larger projects that we are investigating.

1.  We plan to convert all our scanned articles into electronic text
through Optical Character Recognition (OCR).  In order to be able to
search full text articles we will need to develop new search
algorithms.  Our current search system depends on the abstracts being
of fairly uniform length.  This caused some problems at one time when
we included sets of data with abstracts that were 4-5 times as long as
our regular abstracts.  These long abstracts would be found
disproportionally often in searches with many query words (for
instance in query feedback searches), since they generally matched
more words.  We had to reduce the sizes of these abstracts in order to
make the searching work consistently with these data sets.  OCR'd
full text will require new search algorithms to make the search
work at all.  We currently estimate that the implementation of the
full text search capability will take at least 2 years.

2.  The scanned articles frequently contain plots of data.  For most
of these plots the numerical data are not available.  We plan to
develop an interface that will allow our users to select a plot,
display it at high resolution, and digitize the points in the plot by
clicking on them.  This will allow our users to convert plots into
data tables with as much precision as is available on the printed
pages.  At this time we do not know how long it will take us to
implement this new capability.

\subsection{Article Access}

We are currently investigating several new data compression schemes
that would considerably reduce the size of our scanned articles.  This
could considerably improve the utility of the ADS Article Service,
especially on slow links.  We plan to be quite conservative in our
approach to this, since we do not want to be locked into proprietary
compression algorithms that could be expensive to utilize.  At this
point we do not have any time frame in which this might be
accomplished.

\section{Summary}

The ADS Abstract Service has been instrumental in changing the way
astronomers search and access the literature.  One of the reasons for
the immediate acceptance of the ADS by the astronomical community when
the WWW based version became available was the ease of use of the
interface.  It provided access to many advanced features, while still
making it extremely easy to execute simple queries.  Most of the use
is for simple queries, but a significant number of queries utilize
the more sophisticated capabilities.

The search engine that provides this access was crucial to the success
of the ADS as well.  The most important aspects of a search engine are
its speed and its flexibility to accommodate special features.  The
custom-designed software of the ADS search engine proved that at least
in some instances a custom design has considerable advantage over a
general purpose system.  The search speed that we were able to achieve
and the flexibility of the custom design that allows us to quickly
adapt to our users' needs have justified the efforts of developing a
system from scratch that is tailored to the specific data set.

The ADS had 31,533 users issue 780,711 queries and retrieve 15,238,646
references, 615,181 abstracts, and 1,119,122 scanned pages in
November, 1999.  Since the start of the ADS we have served 311,594.261
references and 17,146,370 scanned article pages.  These numbers speak
for them self about the success of the ADS.

\section{Acknowledgment}

Funding for this project has been provided by NASA under NASA Grant
NCC5-189.

\end{document}